\documentclass[%
preprint,
superscriptaddress,
nofootinbib,
showkeys,
amsmath, amssymb,
aps,
pra,
]{revtex4-2}
\usepackage[utf8]{inputenc}
\usepackage[T1]{fontenc}

\usepackage[greek, brazil, english]{babel}
\usepackage{amsmath, amsfonts, amssymb, mathtools, bm}
\usepackage{indentfirst}
\usepackage{lmodern}
\usepackage{fonttable}
\usepackage{alphabeta}
\usepackage{graphicx}
\usepackage{subfigure}
\usepackage{hyperref}
\hypersetup{
    colorlinks = true,
    linkcolor  = blue,
    citecolor  = blue,           % color of links to bibliography
    filecolor  = magenta,
    urlcolor   = blue,
}
\usepackage{calc}
\usepackage{ifthen}
\usepackage{siunitx}
\usepackage{setspace}
\usepackage{color, transparent}	% Controle das cores e transparência

\definecolor{dgreen}{RGB}{0, 153, 0}

\newcommand{\hor}{\mathrm{hor}}
\newcommand{\adj}{\mathrm{adj}}

\renewcommand{\Im}{\textrm{Im}}
\newcommand{\Max}{\textrm{Max}}
\newcommand{\bvec}[1]{\ensuremath{\bm{#1}}}
\renewcommand{\v}[1]{\bvec{#1}}

\let\mgt\gg % from "much greater than"
\renewcommand{\gg}{\bvec{g}}

\let\mlt\ll % from "much less than"
\renewcommand{\ll}{\bvec{\ell}}

\newcommand{\qq}{\bvec{q}}

\newcommand{\xx}{\bvec{x}}

\renewcommand{\cal}[1]{\ensuremath{\mathcal{#1}}}

\newcommand{\calB}{\cal{B}}

\newcommand{\calL}{\cal{L}}

\newcommand{\PAR}[2][-1]{
    \ifthenelse
    { \equal{#1}{-1} }
    { \ensuremath{ {\left( #2 \right) } } }
    { \ensuremath{
            {               \left( \rule{0pt}{ #1 ex} \right. \hspace{-1pt} }
            #2
            { \hspace{-1pt} \left. \rule{0pt}{ #1 ex} \right)               }
} } }
\newcommand{\COL}[2][-1]{
    \ifthenelse
    { \equal{#1}{-1} }
    { \ensuremath{ {\left[\, #2 \,\right] } } }
    { \ensuremath{
            {               \left[ \rule{0pt}{ #1 ex} \right. \hspace{-1pt} }
            \,#2\,
            { \hspace{-1pt} \left. \rule{0pt}{ #1 ex} \right]               }
} } }
\newcommand{\CHA}[2][-1]{
    \ifthenelse
    { \equal{#1}{-1} }
    { \ensuremath{ {\left\{ #2 \right\} } } }
    { \ensuremath{
            {               \left\{ \rule{0pt}{ #1 ex} \right. \hspace{-1pt} }
            #2
            { \hspace{-1pt} \left. \rule{0pt}{ #1 ex} \right\}               }
} } }

\newcommand{\abs}[2][-1]{
    \ifthenelse
    { \equal{#1}{-1} }
    { \ensuremath{ {\left| #2 \right| } } }
    { \ensuremath{
            {               \left| \rule{0pt}{ #1 ex} \right. \hspace{-1pt} }
            #2
            { \hspace{-1pt} \left. \rule{0pt}{ #1 ex} \right|               }
} } }
\DeclareUnicodeCharacter{00B1}{\pm}           % ±
\DeclareUnicodeCharacter{00B9}{^1}            % ¹
\DeclareUnicodeCharacter{00B2}{^{2}}          % ²
\DeclareUnicodeCharacter{00B3}{^3}            % ³
\DeclareUnicodeCharacter{1E0F}{\mathrm{d}}    % ḏ
\DeclareUnicodeCharacter{FF44}{\mathrm{d}}    % ｄ
\DeclareUnicodeCharacter{2202}{\partial}      % ∂
\DeclareUnicodeCharacter{1D735}{\nabla}       % 𝜵
\newcommand{\complexi}{\hspace{.15ex}\textrm{i}\hspace{.15ex}}
\DeclareUnicodeCharacter{012B}{\complexi}     % ī
\DeclareUnicodeCharacter{00EE}{\complexi}     % î
\newcommand{\exponentiale}{\hspace{.15ex}\textrm{e}\hspace{.15ex}}
\DeclareUnicodeCharacter{0113}{\exponentiale} % ē
\DeclareUnicodeCharacter{2260}{\neq}          % ≠
\DeclareUnicodeCharacter{2264}{\leq}   	      % ≤
\DeclareUnicodeCharacter{2265}{\geq}          % ≥
\DeclareUnicodeCharacter{00D7}{\times}        % ×
\DeclareUnicodeCharacter{00B7}{\cdot}         % ·
\DeclareUnicodeCharacter{222B}{\int}          % ∫
\DeclareUnicodeCharacter{25A1}{\square}       % □
\DeclareUnicodeCharacter{2113}{\ell}          % ℓ
\DeclareUnicodeCharacter{0127}{\hbar}         % ħ
\DeclareUnicodeCharacter{221E}{\infty}        % ∞
\DeclareUnicodeCharacter{2020}{^\dagger\!}    % †
\DeclareUnicodeCharacter{2021}{^\ddagger\!}   % ‡
\DeclareUnicodeCharacter{22EF}{\cdots}        % ⋯
\DeclareUnicodeCharacter{00A0}{{}}
\DeclareUnicodeCharacter{2070}{^0}     % ⁰
\DeclareUnicodeCharacter{00B9}{^1}     % ¹
\DeclareUnicodeCharacter{00B2}{^2}     % ²
\DeclareUnicodeCharacter{00B3}{^3}     % ³
\DeclareUnicodeCharacter{2074}{^4}     % ⁴
\DeclareUnicodeCharacter{2075}{^5}     % ⁵
\DeclareUnicodeCharacter{2076}{^6}     % ⁶
\DeclareUnicodeCharacter{2077}{^7}     % ⁷
\DeclareUnicodeCharacter{2078}{^8}     % ⁸
\DeclareUnicodeCharacter{2079}{^9}     % ⁹
\DeclareUnicodeCharacter{207A}{^{+}}   % ⁺
\DeclareUnicodeCharacter{207B}{^{-}}   % ⁻
\DeclareUnicodeCharacter{207C}{^{=}}   % ⁼
\DeclareUnicodeCharacter{2080}{_0}     % ₀
\DeclareUnicodeCharacter{2081}{_1}     % ₁
\DeclareUnicodeCharacter{2082}{_2}     % ₂
\DeclareUnicodeCharacter{2083}{_3}     % ₃
\DeclareUnicodeCharacter{2084}{_4}     % ₄
\DeclareUnicodeCharacter{2085}{_5}     % ₅
\DeclareUnicodeCharacter{2086}{_6}     % ₆
\DeclareUnicodeCharacter{2087}{_7}     % ₇
\DeclareUnicodeCharacter{2088}{_8}     % ₈
\DeclareUnicodeCharacter{2089}{_9}     % ₉
\DeclareUnicodeCharacter{208A}{_{\!+}} % ₊
\DeclareUnicodeCharacter{208B}{_{\!-}} % ₋
\DeclareUnicodeCharacter{208C}{_{\!=}} % ₌
\DeclareUnicodeCharacter{2098}{_m}     % ₘ
\DeclareUnicodeCharacter{2099}{_n}     % ₙ
\DeclareUnicodeCharacter{2261}{\equiv}          % ≡
\DeclareUnicodeCharacter{2243}{\simeq}          % ≃
\DeclareUnicodeCharacter{223C}{\sim}            % ∼
\DeclareUnicodeCharacter{21D2}{\Rightarrow}     % ⇒
\DeclareUnicodeCharacter{27F9}{\implies}        % ⟹
\DeclareUnicodeCharacter{2192}{\rightarrow}     %  →
\DeclareUnicodeCharacter{219B}{\nrightarrow}    % ↛
\DeclareUnicodeCharacter{27F6}{\longrightarrow} % ⟶
\expandafter\newcommand\csname u8:\detokenize{√}\endcsname[2][]{%
    \ifthenelse
    { \equal{#1}{} }
    { \sqrt{#2} }
    { \sqrt[#1]{#2} }
}

\makeatletter
\newcommand*{\msh@ifnextchar}[3]{%
    \def\msh@temp{\msh@@ifnextchar{#1}{#2}{#3}}%
    \futurelet\msh@token\msh@temp
}
\newcommand*{\msh@@ifnextchar}[1]{%
    \ifx\msh@token#1%
    \expandafter\@firstoftwo
    \else
    \expandafter\@secondoftwo
    \fi
}

\mathchardef\msh@code@less=\mathcode`\<\relax
\mathchardef\msh@code@greater=\mathcode`\>\relax
\mathchardef\msh@code@minus=\mathcode`\-\relax
\mathchardef\msh@code@plus=\mathcode`\+\relax
\mathchardef\msh@code@equal=\mathcode`\=\relax

\@ifdefinable{\resetmathshorthands}{%
    \edef\resetmathshorthands{%
        \mathcode\number`\<=\msh@code@less
        \mathcode\number`\>=\msh@code@greater
        \mathcode\number`\-=\msh@code@minus
        \mathcode\number`\+=\msh@code@plus
        \mathcode\number`\==\msh@code@equal
    }%
}

\begingroup
\catcode`\<=\active
\catcode`\>=\active
\catcode`\-=\active
\catcode`\+=\active
\catcode`\==\active
\edef={\string=}%
\@ifdefinable{\setmathshorthands}{%
    \xdef\setmathshorthands{%
        \mathcode\number`\<="8000 %
        \mathcode\number`\>="8000 %
        \mathcode\number`\-="8000 %
        \mathcode\number`\+="8000 %
        \mathcode\number`\=="8000 %
        \let\noexpand<\noexpand\msh@less
        \let\noexpand>\noexpand\msh@greater
        \let\noexpand-\noexpand\msh@minus
        \let\noexpand+\noexpand\msh@plus
        \let\noexpand=\noexpand\msh@equal
    }%
}%
\endgroup

\newcommand*{\msh@less}{%
    \msh@ifnextchar<{%
        \mlt    \@gobble
    }{%
        \msh@ifnextchar>{%
            \neq\@gobble
        }{%
            \msh@ifnextchar={%
                \expandafter\msh@less@equal\@gobble
            }{%
                \msh@ifnextchar-{%
                    \expandafter\msh@less@minus\@gobble
                }{%
                    \msh@code@less
                }%
            }%
        }%
    }%
}
\newcommand*{\msh@less@equal}{%
    \msh@ifnextchar={%
        \Leftarrow\@gobble
    }{%
        \msh@ifnextchar>{%
            \Leftrightarrow\@gobble
        }{%
            \leq
        }%
    }%
}
\newcommand*{\msh@less@minus}{%
    \msh@ifnextchar-{%
        \leftarrow\@gobble
    }{%
        \msh@ifnextchar>{%
            \leftrightarrow\@gobble
        }{%
            \msh@code@less\msh@code@minus
        }%
    }%
}

\newcommand*{\msh@greater}{%
    \msh@ifnextchar>{%
        \mgt\@gobble
    }{%
        \msh@ifnextchar={%
            \geq\@gobble
        }{%
            \msh@code@greater
        }%
    }%
}

\newcommand*{\msh@minus}{%
    \msh@ifnextchar-{%
        \expandafter\msh@minus@minus\@gobble
    }{%
        \msh@ifnextchar+{%
            \mp\@gobble
        }{%
            \msh@code@minus
        }%
    }%
}
\newcommand*{\msh@minus@minus}{%
    \msh@ifnextchar>{%
        \rightarrow\@gobble
    }{%
        \msh@code@minus\msh@code@minus
    }%
}

\newcommand*{\msh@plus}{%
    \msh@ifnextchar-{%
        \pm\@gobble
    }{%
        \msh@code@plus
    }%
}

\newcommand*{\msh@equal}{%
    \msh@ifnextchar={%
        \expandafter\msh@equal@equal\@gobble
    }{%
        \msh@code@equal
    }%
}
\newcommand*{\msh@equal@equal}{%
    \msh@ifnextchar>{%
        \Rightarrow\@gobble
    }{%
        \equiv
    }%
}
\makeatother

\everymath{\setmathshorthands}
\everydisplay{\setmathshorthands}
\let\originalleft\left
\let\originalright\right
\renewcommand{\left}{\mathopen{}\mathclose\bgroup\originalleft}
\renewcommand{\right}{\aftergroup\egroup\originalright}

\DeclareSymbolFont{mylargesymbols}{OMX}{ccex}{m}{n}
\DeclareMathDelimiter{\lbrace}{\mathopen}{symbols}{"66}{mylargesymbols}{"08}
\DeclareMathDelimiter{\rbrace}{\mathclose}{symbols}{"67}{mylargesymbols}{"09}
\DeclareMathDelimiter{(}{\mathopen}{operators}{"28}{mylargesymbols}{"00}
\DeclareMathDelimiter{)}{\mathclose}{operators}{"29}{mylargesymbols}{"01}
\DeclareMathDelimiter{[}{\mathopen}{operators}{"5B}{mylargesymbols}{"02}
\DeclareMathDelimiter{]}{\mathclose}{operators}{"5D}{mylargesymbols}{"03}
\DeclareMathSymbol{\braceld}{\mathord}{mylargesymbols}{"7A}
\DeclareMathSymbol{\bracerd}{\mathord}{mylargesymbols}{"7B}
\DeclareMathSymbol{\bracelu}{\mathord}{mylargesymbols}{"7C}
\DeclareMathSymbol{\braceru}{\mathord}{mylargesymbols}{"7D}

\expandafter\newcommand\csname u8:\detokenize{⦅}\endcsname[1][]{%
    \ifthenelse
    { \equal{#1}{} }
    { \left( }
    { \left( \rule{0pt}{ #1 ex * 17/30 + 26ex/30 } \right. \hspace{-1pt} }
}
\expandafter\newcommand\csname u8:\detokenize{⦆}\endcsname[1][]{%
    \ifthenelse
    { \equal{#1}{} }
    { \right) }
    { \hspace{-1pt} \left. \rule{0pt}{ #1 ex * 17/30 + 26ex/30 } \right) }
}
\expandafter\newcommand\csname u8:\detokenize{⟦}\endcsname[1][]{%
    \ifthenelse
    { \equal{#1}{} }
    { \left[ }
    { \left[ \rule{0pt}{ #1 ex * 17/30 + 26ex/30 } \right. \hspace{-2pt} }
}
\expandafter\newcommand\csname u8:\detokenize{⟧}\endcsname[1][]{%
    \ifthenelse
    { \equal{#1}{} }
    { \right] }
    { \left. \rule{0pt}{ #1 ex * 17/30 + 26ex/30 } \right] }
}
\expandafter\newcommand\csname u8:\detokenize{⦃}\endcsname[1][]{%
    \ifthenelse
    { \equal{#1}{} }
    { \left\{ }
    { \left\{ \rule{0pt}{ #1 ex * 17/30 + 26ex/30 } \right. \hspace{-1pt} }
}
\expandafter\newcommand\csname u8:\detokenize{⦄}\endcsname[1][]{%
    \ifthenelse
    { \equal{#1}{} }
    { \right\} }
    { \hspace{-1pt} \left. \rule{0pt}{ #1 ex * 17/30 + 26ex/30 } \right\} }
}
\expandafter\renewcommand\csname u8:\detokenize{⟨}\endcsname[1][]{%
    \ifthenelse
    { \equal{#1}{} }
    { \langle }
    { \left\langle \rule{0pt}{ #1 ex * 17/30 + 26ex/30 } \right. \hspace{-1pt} }
}
\expandafter\renewcommand\csname u8:\detokenize{⟩}\endcsname[1][]{%
    \ifthenelse
    { \equal{#1}{} }
    { \rangle }
    { \hspace{-1pt} \left. \rule{0pt}{ #1 ex * 17/30 + 26ex/30 } \right\rangle }
}
\begin{document}

\title{Configuration entropy and stability of bottomonium radial excitations in a plasma with magnetic fields}

\author{Nelson R. F. Braga}
\email{braga@if.ufrj.br}
\affiliation{Instituto de Física, Universidade Federal do Rio de Janeiro, Caixa Postal 68528, RJ 21941-972, Brazil.}

\author{Yan F. Ferreira}
\email{yancarloff@pos.if.ufrj.br}
\affiliation{Instituto de Física, Universidade Federal do Rio de Janeiro, Caixa Postal 68528, RJ 21941-972, Brazil.}

\author{Luiz F. Ferreira}
\email{luiz.faulhaber@ufabc.edu.br}
\affiliation{CCNH, Universidade Federal do ABC –- UFABC, 09210-580, Santo André, Brazil.\rule[-30pt]{0pt}{1pt}}

\begin{abstract}
\setstretch{1.2}

\noindent
\hspace{.75cm}\!\!
Heavy vector mesons produced in a heavy ion collision are important sources of information about the quark gluon plasma (QGP). For instance, the fraction of bottomonium states observed in such a collision is altered by the dissociation effect caused by the plasma. So, it is very important to understand how the properties of the plasma, like temperature $(T)$, density and the presence of background magnetic fields $(eB)$, affect the dissociation of bottomonium in the thermal medium. AdS/QCD holographic models provide a tool for investigating the properties of heavy mesons inside a thermal medium. The meson states are represented by quasinormal modes in a black hole geometry. In this work we calculate the quasinormal modes and the associated complex frequencies for the four lowest levels of radial excitation of bottomonium inside a plasma with a magnetic field background. We also calculate the differential configuration entropy (DCE) for all these states and investigate how the dissociation effect produced by the magnetic field is translated into a dependence of the DCE on $eB$. An interesting result obtained in this study is that the DCE increases with the radial excitation level $n$. Also, a nontrivial finding of this work is that the energy density associated with the bottomonium quasinormal modes presents a singularity near the black hole horizon for some combination of values of $T, eB$ and $n$. As we show here, it is possible to separate the singular factor and define a square integrable quantity that provides a DCE that is always finite. In addition, we discovered that, working with the potentially singular energy density, one finds a very interesting way to use the DCE as a tool for determining the dissociation temperature of the meson quasisates.
\\[-.5\baselineskip]

\end{abstract}


\maketitle
\newpage

% ----------------------------------------------------------
\vspace{.75\baselineskip}
\section{Introduction}

One of the most interesting challenges faced by physicists in the present days is to understand the properties of the quark gluon plasma (QGP). This state of matter, in which quarks and gluons interact strongly but are are not confined into hadrons, is formed in heavy ion collisions. For reviews about the QGP, see for example \cite{Bass:1998vz, Scherer:1999qq, Shuryak:2008eq, Casalderrey-Solana:2011dxg}. The QGP lives for a very short time and the available information about it comes from the particles reaching the detectors after hadronization. Among those particles, bottomonium vector mesons, which are composed of a bottom-antibottom quark pair are particularly important \cite{Andronic:2015wma, Rothkopf:2019ipj}. These particles are created in the collision and then partially dissociate in the medium when the QGP is formed. The fraction of bottomonium final states observed in a heavy ion collision depends on how strong is the dissociation effect caused by the plasma. On the other hand, thermal dissociation depends on the properties of the plasma, such as: temperature, density and the presence of magnetic fields. That is why bottomonium can serve as an important probe of QGP properties.

The possibility of dissociation in the medium corresponds to a form of instability of bottomonium states. A very interesting tool to study stability of physical systems is the configuration entropy (CE). In recent years many examples appeared in the literature, involving various kinds of physical systems, where an increase in the CE is associated with an increase in the instability of the system. For example, the authors of \cite{Gleiser:2011di, Gleiser:2012tu, Gleiser:2013mga, Stephens:2019tav} have analyzed the informational entropy in compact astrophysical objects and the field theory. In the setup of AdS/QCD, the differential configuration entropy (DCE) has been applied to study new features of heavy-quark mesons, baryons, tensorial mesons, pomerons, odderons, ligth-flavor mesons and the deconfinement/confinement behavior of the hard-wall and soft-wall models \cite{Bernardini:2016qit, Braga:2017fsb, Braga:2018fyc, Ferreira:2019inu, Colangelo:2018mrt, Ferreira:2019nkz, Braga:2020myi, Ferreira:2020iry, MarinhoRodrigues:2020ssq, Braga:2020opg, Braga:2021zyi}. Other applications of the DCE in quantum chromodynamics (QCD), condensed matter physics, cosmology, AdS/CFT and heavy ion collisions can be found in references \cite{Zhao:2019xle, Karapetyan:2016fai, Karapetyan:2017edu, Karapetyan:2018oye, Karapetyan:2018yhm, Bazeia:2018uyg, Karapetyan:2019ran, Karapetyan:2020yhs, Alves:2020cmr, Bazeia:2021stz, Lee:2018zmp, Bazeia:2018uyg, Ma:2018wtw, Correa:2015vka, Braga:2019jqg, Braga:2016wzx}.

The gauge theory in the AdS/CFT correspondence is conformal and the radial AdS coordinate is interpreted as the holographic energy scale of the gauge theory. There are no energy parameters in the AdS/CFT correspondence. In order to describe the strong interactions, one can build up phenomenological AdS/QCD models where conformal symmetry is broken. This can be done in different ways. In the holographic model that we will consider here this is done by introducing a background scalar field in the action integral. As will be shown in section \ref{sec: model}, this field depends on three energy parameters.

In this work we use a holographic AdS/QCD model in order to represent bottomonium states in a plasma. We calculate the quasinormal modes, the associated complex frequencies and the corresponding CE for four different levels of radial excitation of bottomonium quasistates inside a plasma with a magnetic field background. Then we investigate how the instability, corresponding in this case to the dissociation in the thermal medium, is translated into a dependence of the configuration entropy on the field.

The CE is a continuous version of the information entropy introduced by Shannon \cite{Shannon:1948zz}:
\begin{gather}
    S_{\text{Shannon}}^{} = - \sum_{n} p_n \ln p_n.
\end{gather}
This quantity represents the amount of information contained in a variable $x$ that can assume discrete values $x_n$, each of them with a probability $p_n$. Inspired by this definition, one introduces the configuration entropy \cite{Gleiser:2013mga} in an analogous way, but for continuous variables:
\begin{gather}
    S = -∫ｄ^d\vec{k}\, ϵ(\vec{k}) \ln ϵ(\vec{k}),
\shortintertext{where}
    ϵ(\vec{k}) = \dfrac{ |\tilde{R}(\vec{k})|^2 }{\displaystyle \Max(|\tilde{R}(\vec{k})|^2) \rule{0ex}{2.4ex}},
    \label{eq: intro - modal fraction}
\end{gather}
is called the modal fraction. It is defined using the momentum space Fourier transform of a normalizable (square integrable) function in coordinate space $R(\vec{r})$:
\begin{gather}
    \tilde{R}(\vec{k}) = ∫ｄ^d\vec{r}\, R(\vec{r})\,  \mathrm{e}^{-î \vec{k}\cdot\vec{r}}.
    \label{eq: intro - R(k)}
\end{gather}
The energy density $ρ(\vec{r})$ of the physical system is usually taken as the normalizable function $R(\vec{r})$ which defines the modal fraction. One finds in the literature an alternative definition for the modal fraction, where in equation (\ref{eq: intro - modal fraction}) one uses the normalization integral of the square of $\tilde{R}(\vec{k})$ in the denominator, instead of the maximum value. However, the alternative definition of the modal fraction can lead to negative values for the configuration entropy, as pointed out in reference \cite{Stephens:2019tav}. In contrast, the definition of equation (\ref{eq: intro - modal fraction}) for the modal fraction ensures the positivity of the configuration entropy for continuous variables. For this reason we will use the definition of equation (\ref{eq: intro - modal fraction}), which is called differential configuration entropy (DCE) in this work.

In reference \cite{Braga:2020hhs}, the authors used holography to study the dissociation of the ground state of charmonium in a plasma as function of the magnetic field $B$ and calculated the corresponding configuration entropy. Here we consider bottomonium, which undergoes dissociation at higher temperatures, and study the dissociation of the different radial excitation levels, not only the ground state. This way, we will be able to analyze the dependence of the configuration entropy on the excitation level, which was not analyzed in this previous work. The lowest radial excitations of bottomonium survive the deconfinement transition and therefore are also important probes of the plasma properties.

This article is organized this way: in section \ref{sec: model} we review the holographic AdS/QCD bottom-up model that describes bottomonium in a plasma. In section \ref{sec: QNMs} we obtain the quasinormal modes as functions of the magnetic field $B$. In section \ref{sec: energy density} we use the field that represents the quasiparticles with corresponding quasinormal frequency to calculate the energy density of the system. Then, in section \ref{sec: CE}, we use the results obtained in order to find the configuration entropy of the system. Then, in section \ref{sec: NRCE}, we reexamine the calculation of the DCE, considering the nonregularized form of the energy density, and find a way to characterize the complete dissociation in the medium. Finally, in section \ref{sec: Conclusions}, we present a discussion about the results obtained.

% ----------------------------------------------------------
\vspace{.75\baselineskip}
\section{Holographic description of bottomonium in a plasma with magnetic fields}
\label{sec: model}

The study of the kind of holographic model for heavy vector mesons that we will consider started in references \cite{Braga:2015jca, Braga:2016wkm, Braga:2017oqw}. The model used in references \cite{Braga:2015jca, Braga:2016wkm, Braga:2017oqw} for quarkonium includes one energy parameter associated with the mass spectrum and another associated with the decay process. However, the decay constants obtained for $J/ψ$ from this model are $40\%$ smaller than the experimental result. Then an improved version, which we will use here, was developed in references \cite{Braga:2017bml, Braga:2018zlu, Braga:2018hjt, Braga:2019yeh} considering three energy parameters: one associated with the quark mass, other associated with the string tension and the third one associated with the nonhadronic decay of the mesons. This model provides good estimates not only for the quarkonium masses, but also for the decay constants. This is important since we are concerned with the study of bottomonium in the plasma. The decay constants are related to the heights of the peaks of the spectral functions. It is, therefore, crucial to have an accurate fit for these quantities in order to describe quarkonium in a thermal medium \cite{Braga:2017bml}. For other approaches to the holographic description of bottomonium see, for example \cite{MartinContreras:2021bis, Zollner:2021stb}.

Vector mesons are represented by a 5-dimensional dual vector field $V_m$ with an action integral of the form\footnote{Note that we are not studying the effect the magnetic field makes over the bottomonium itself, but over the background plasma. As consequence of it, the metric we use depends on $B$ and the Lagrangian does not.}
\begin{gather}
    I = -\dfrac{1}{4 g_5^2}∫ｄ^4x ｄz\, √{-g}\, \calL,
\end{gather}
with the Lagrangian density $\calL = e^{-ϕ(z)} g^{mp} g^{nq} F_{mn}^* F_{pq}$, where $F_{mn} = \nabla_{\!m} V_n - \nabla_{\!n} V_m = ∂_m V_n - ∂_n V_m$ with $\nabla_{\!m}$ being the covariant derivative. Notice that the Lagrangian is unchanged by the gauge transformation $V_μ \rightarrow V_μ - ∂_μ Λ$, where $Λ$ is an arbitrary function. We take the magnetic field in the $x^3$ direction and use the perturbative metric in $eB$ \cite{Dudal:2015wfn}. Note that in the bottom up model studied in reference \cite{Dudal:2015wfn}, the relation between the physical magnetic field, $eB$, and the magnetic field in their action, $\calB$, was found by matching the action of gauge fields in the soft wall model normalized by QCD flavor-flavor correlators and the Maxwell action normalized in \cite{DHoker:2009ixq}. This provides the relation between the physical magnetic field and the bulk five-dimensional magnetic field, $eB = 1.6\,\calB$. For this reason, the factor $1.6^2$ will appear in the metric functions. For more details see the Appendix in reference \cite{Dudal:2015wfn}.
\begin{gather}
    ｄs^2 = \dfrac{R^2}{z^2} \CHA{-f(z)ｄt^2 + d(z)\COL{(ｄx^1)^2 +(ｄx^2)^2} + h(z)(ｄx^3)^2 + \dfrac{1}{f(z)}ｄz^2},
\end{gather}
where\footnote{One can find a primitive of $(y^3 \ln y)/(1-y^4)$ in terms of the polylogarithm function.}\vspace{-.5\baselineskip}
\begin{align}\label{c1}
    f(z) &= 1 - \dfrac{z^4}{z_h^4} + \dfrac{2}{3}\dfrac{e^2 B^2}{1.6^2}z^4\ln\dfrac{z}{z_h}, \\
    h(z) &= 1 + \dfrac{8}{3}\dfrac{e^2 B^2}{1.6^2} z_h^4 ∫_{0}^{z/z_h}\dfrac{y^3 \ln y}{1-y^4}ｄy, \\
    d(z) &= 1 - \dfrac{4}{3}\dfrac{e^2 B^2}{1.6^2} z_h^4 ∫_{0}^{z/z_h}\dfrac{y^3 \ln y}{1-y^4}ｄy,
\end{align}
$R$ is the AdS radius and $z_h$ is the horizon position\footnote{The authors in reference \cite{Dudal:2015wfn} have introduced an extra length parameter $l_d$ in $f(z)$. Independently of the choice of $l_d$, we have a solution of the Einstein's equation as discussed in \cite{Dudal:2015wfn}. The physical quantities are independent of $l_d$ because this parameter can be dropped out. More details can be found in \cite{Dudal:2015wfn}.}. The plasma temperature is given by \cite{Hawking:1974rv, Hawking:1975vcx}
\begin{gather}
    T = \dfrac{|f'(z_h)|}{4π} = \dfrac{1}{4π}\abs{\dfrac{4}{z_h} - \dfrac{2}{3}\dfrac{e^2 B^2}{1.6^2}}.
    \label{eq: temperature}
\end{gather}
In the absence of the plasma ($z_h \rightarrow ∞$) and of a magnetic field ($B \rightarrow 0$), that means, in the vacuum, the space is just a 5-dimensional anti-de Sitter one. The background field $ϕ(z)$ used is
\begin{gather}
    ϕ(z) = κ^2 z^2 + Mz + \tanh\!\PAR{\dfrac{1}{Mz} - \dfrac{κ}{√{Γ}}}.
\end{gather}
We take the masses and decay constants of bottomonium in the vacuum from the particle data group table \cite{ParticleDataGroup:2020ssz} and the parameters that best fit them are $κ_b = \SI{2.45}{\giga\eV}$, $√{Γ_b} = \SI{1.55}{\giga\eV}$ and $M_b = \SI{6.2}{\giga\eV}$.

We now choose a Fourier component of the field, $V_m(t, \xx, z) = η_m v(ω, \qq, z) e^{ + î η_{μν} q^μ x^ν}$, with polarization $η_m$. We choose the gauge $V_z = 0$ and consider the vector meson at rest, so that $\qq = \v{0}$. This gives $V_μ(t, \xx, z) = V_μ(t, z) = η_μ v(ω, z) e^{-îωt}$.

For transverse polarizations, $η_μ = (0,1,0,0)$ and $η_μ = (0,0,1,0)$, we have
\begin{align}
    \dfrac{ω^2}{f(z)^2} v(z) + \PAR{-\frac{1}{z} + \frac{f'(z)}{f(z)} + \dfrac{h'(z)}{2h(z)} - ϕ'(z)} v'(z) + v''(z) &= 0,
    \label{eq: eq motion T}
\end{align}
where the prime stands for the derivative with respect to $z$ and, for simplicity, we use the notation $v(z)$ for $v(ω,z)$, omitting the dependency on $ω$. For longitudinal polarization, $η_μ = (0,0,0,1)$, we have
\begin{align}
    \dfrac{ω^2}{f(z)^2} v(z) + \PAR{-\frac{1}{z} + \frac{f'(z)}{f(z)} + \frac{d\hspace{.75pt}'(z)}{d(z)} - \dfrac{h'(z)}{2h(z)} - ϕ'(z)} v'(z) + v''(z) &= 0.
    \label{eq: eq motion L}
\end{align}
The solutions of equations \eqref{eq: eq motion T} and \eqref{eq: eq motion L}, with appropriate boundary conditions, will provide the quasinormal modes. Note that when $eB = 0$, equations \eqref{eq: eq motion T} and \eqref{eq: eq motion L} are the same and the two states, longitudinal and transversal, are degenerate. The presence of the magnetic field breaks this degeneracy.

% ----------------------------------------------------------
\vspace{.75\baselineskip}
\section{Quasinormal Modes}
\label{sec: QNMs}

Quasinormal modes (QNMs) are field solutions with complex frequencies that satisfy the infalling wave condition on the horizon and represent the meson quasistates. In order to find the QNMs, one needs to analyze the behavior of the field near the horizon. One can expand the coefficients of $v(z)$, $v'(z)$ and $v''(z)$ in \eqref{eq: eq motion T} and \eqref{eq: eq motion L} in powers of $ (z - z_h) $ and keep only the dominant terms. This gives, for longitudinal and transversal polarizations
\begin{gather}
    \dfrac{ω^2}{\displaystyle\rule{0pt}{11.5pt}f'(z_h)^2(z - z_h)^2}\,v_{\hor}^{}(z) + \dfrac{1}{z - z_h}\,v_{\hor}'(z) + \,v_{\hor}''(z) = 0,
    \label{eq: eq of motion near the horizon}
\end{gather}
where we used $f(z_h) = 0$. In terms of the temperature \eqref{eq: temperature}, we can write \eqref{eq: eq of motion near the horizon} as
\begin{gather}
    \dfrac{ω^2}{\displaystyle\rule{0pt}{11.5pt}(4πT)^2}\,v_{\hor}^{}(z) + (z - z_h)\,v_{\hor}'(z) + (z - z_h)^2\,v_{\hor}''(z) = 0,
\end{gather}
whose solutions are
\begin{gather}
    \PAR[3]{ 1 - \dfrac{z}{z_h} }^{\!\!+îω/4πT}
    \qquad\qquad
    \text{and}
    \qquad\qquad
    \PAR[3]{ 1 - \dfrac{z}{z_h} }^{\!\!-îω/4πT}.
\end{gather}
The second solution, with the minus sign in the exponent, corresponds to an infalling wave at the horizon, while the first solution, with positive sign, to an outgoing wave. This becomes clear if one change to the Regge-Wheeler tortoise coordinate \cite{Mamani:2013ssa}. The quasinormal complex frequencies are found by imposing the field to satisfy the condition of being an infalling wave at the horizon, and the Dirichlet condition at the boundary $v(0) = 0$ \cite{Kaminski:2009ce, Janiszewski:2015ura}.

We find the quasinormal field $v(z)$ in terms of $ω$ by solving the complete equations of motion (\ref{eq: eq motion T}) and (\ref{eq: eq motion L}) numerically with the conditions $v(z_0) = v_{\hor,\,p}(z_0)$ and $v'(z_0) = v_{\hor,\,p}'(z_0)$, where $z_0 = z_h - ϵ$, with $ϵ$ small, and $ v_{\hor,\,p}$ is taken in the form
\begin{gather}
    v_{\hor,\,p}(z) = \PAR[3]{ 1 - \dfrac{z}{z_h} }^{\!\!-îω/4πT}
    \sum_{n=0}^{p} a_n \PAR{1 - \dfrac{z}{z_h}}^{\!\!n},
    \label{eq: vhor}
\end{gather}
which is the infalling horizon solution $v_{\hor}$ times a polynomial perturbation introduced in order to calculate the field at $z_0 = z_h - ϵ$. The first coefficient of the polynomial perturbation is $a_0 = 1$ and the other coefficients $a_n$ are determined by substituting \eqref{eq: vhor} into the equation of motion \eqref{eq: eq motion T}, for longitudinal polarization, or \eqref{eq: eq motion L}, for transverse polarization. This is how we impose the infalling wave condition at the horizon in the numerical calculation.

It is important to note that due to the highly oscillatory behavior of the factor $(1 - z/z_h )^{-îω/4πT}$ near $z = z_h$, $ϵ$ cannot be too small \cite{Kaminski:2009ce}. On the other hand, the larger is the value of $ϵ$, the greater must be $p$, the order of the polynomial perturbation in \eqref{eq: vhor}. One has to consider this two factors in order to fix the value of $ϵ$.

Following this procedure we build a parametric numerical solution of the equation of motion with parameter $ω$. With this parametric numerical solution we impose the Dirichlet condition at the boundary by numerically solving the equation $v(ω, z\!=\!0) = 0$ for complex $ω$. This determines the quasinormal frequency. The numeric solution of the equation of motion with the value of $ω$ obtained will then give the field $v$ in the interval from 0 to $z_0$.

\begin{figure}[htb!]
    \centering
    \vspace{\baselineskip}
    \includegraphics[scale=.4]{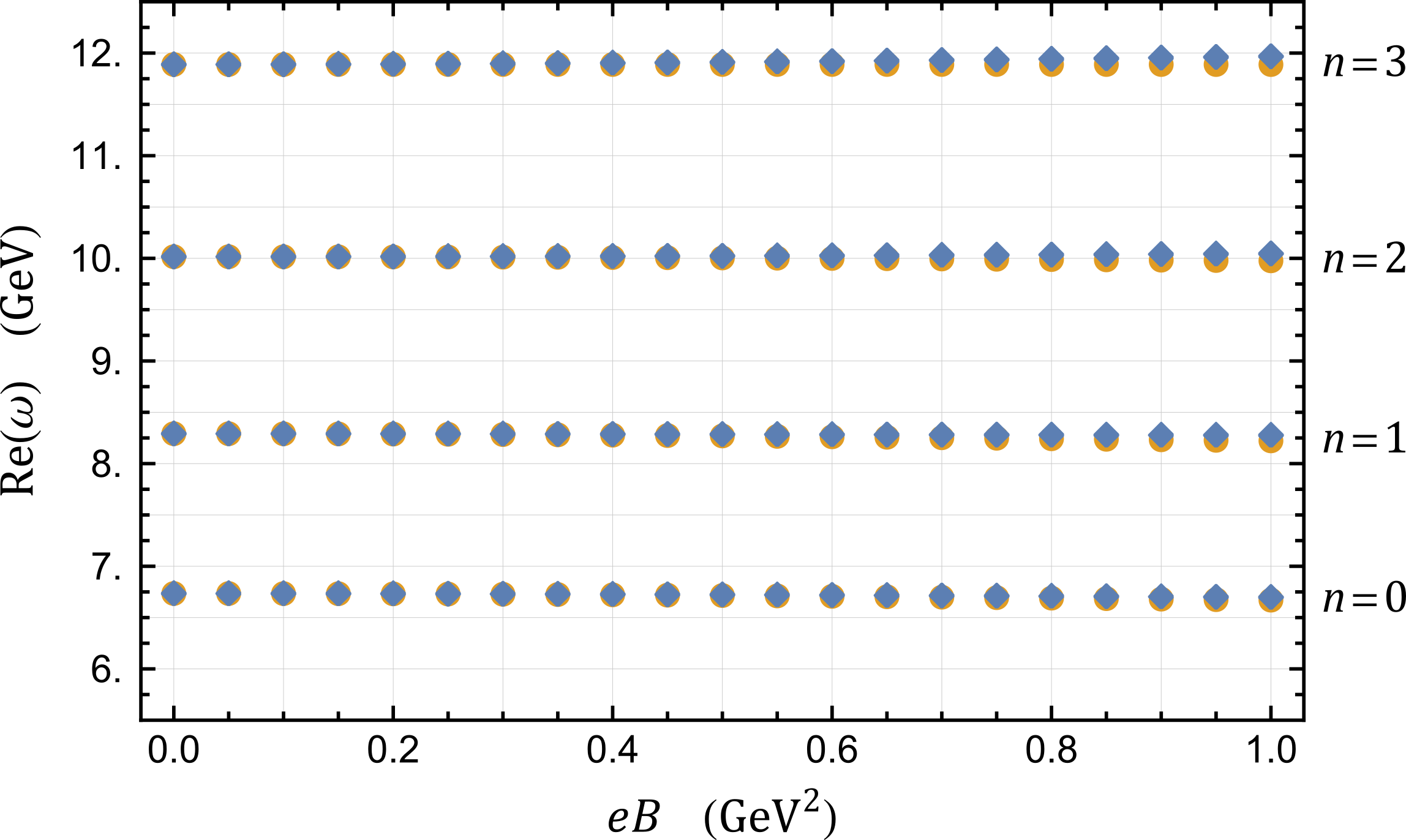} \quad
    \includegraphics[scale=.4]{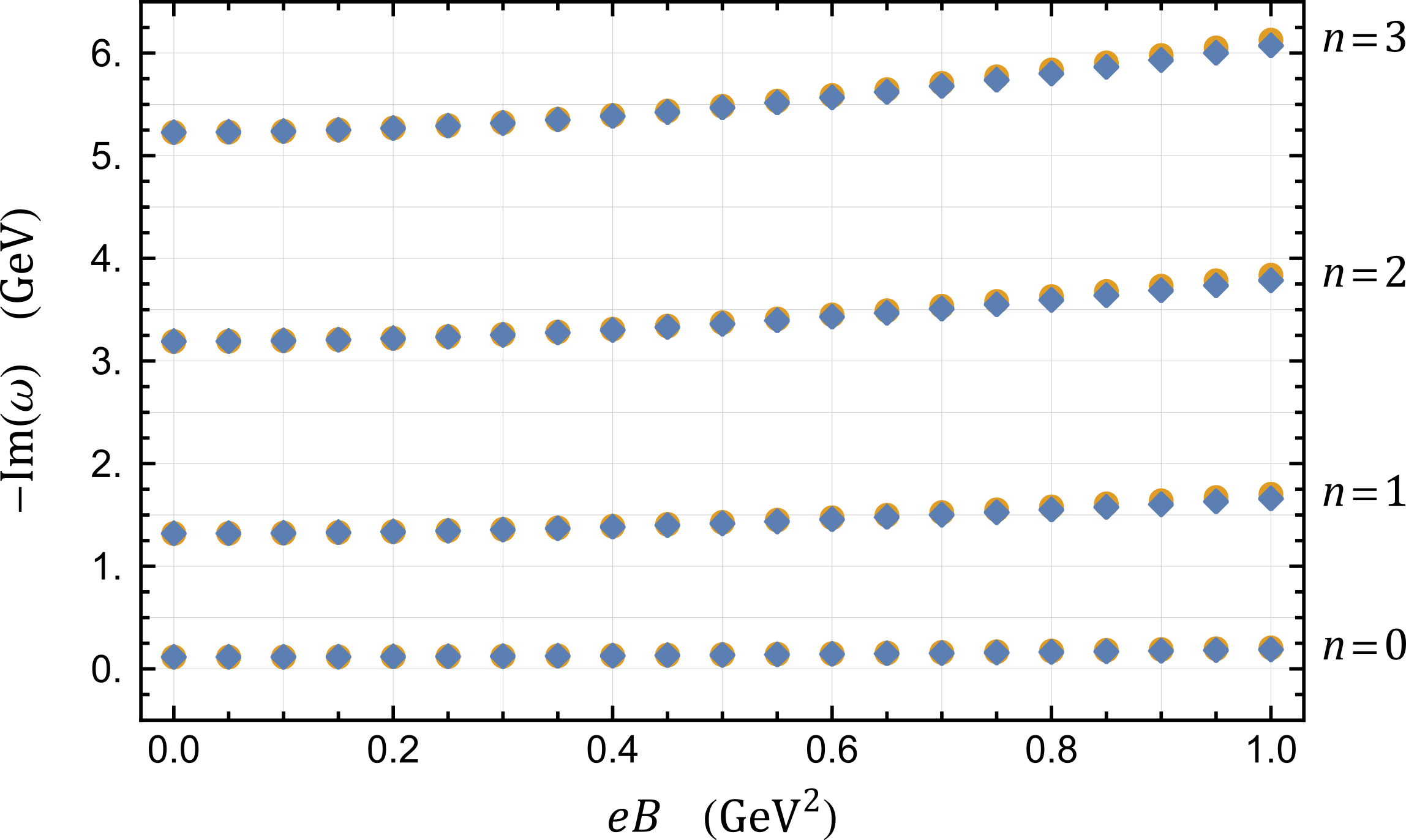} \\[8pt]
    \raisebox{2pt}{\includegraphics[scale=.36]{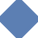}} \quad Transverse Polarization\\[3pt]
    \raisebox{2pt}{\includegraphics[scale=.36]{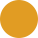}}    \quad Longitudinal Polarization
    \caption{Quasinormal frequencies for transverse and longitudinal polarizations of the different excitation levels as function of the magnetic field $B$ and for temperature fixed at $T = \SI{300}{\mega\eV}$.}
    \label{fig: QNMs1}
\end{figure}

\begin{figure}[htb!]
    \centering
    \includegraphics[scale=.41]{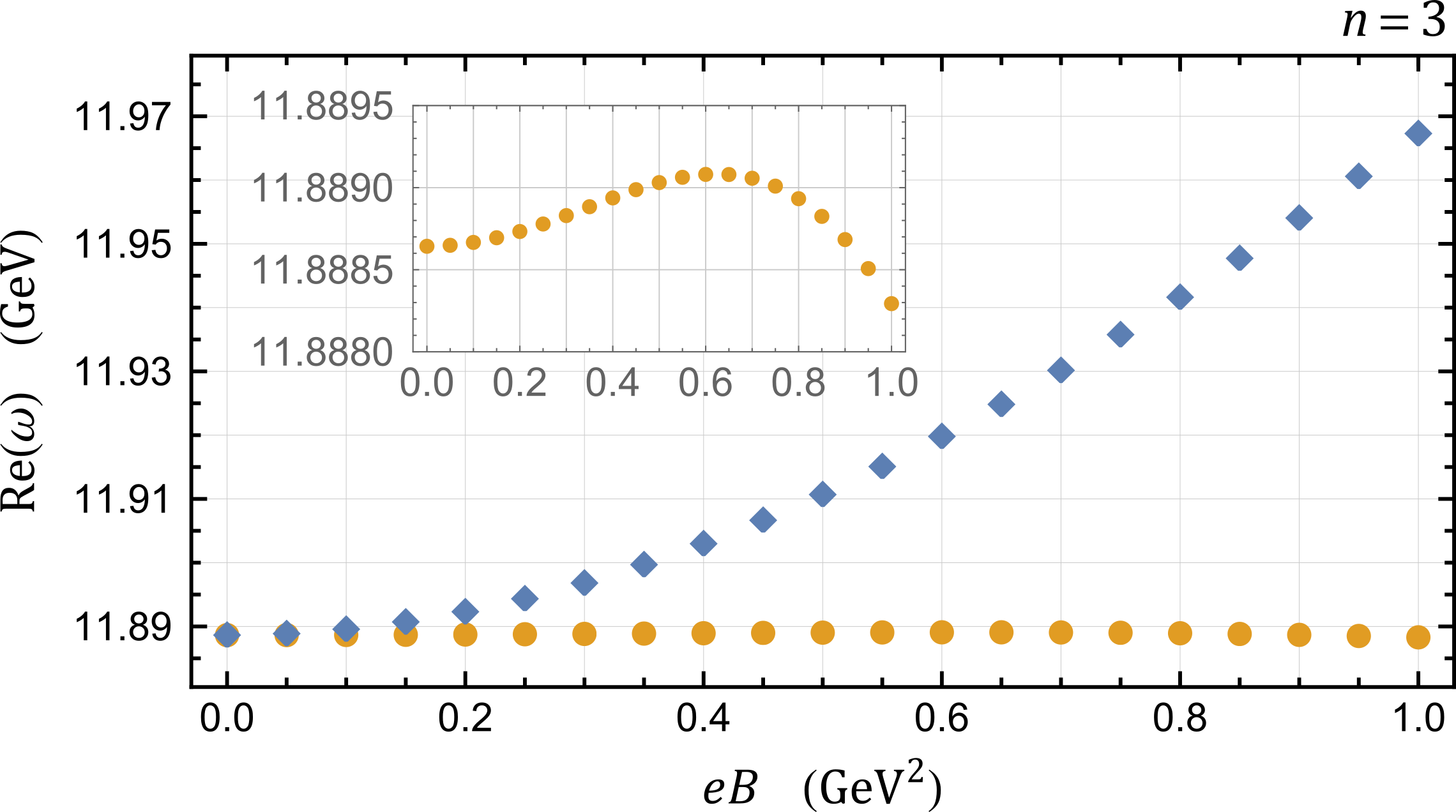} \quad
    \includegraphics[scale=.41]{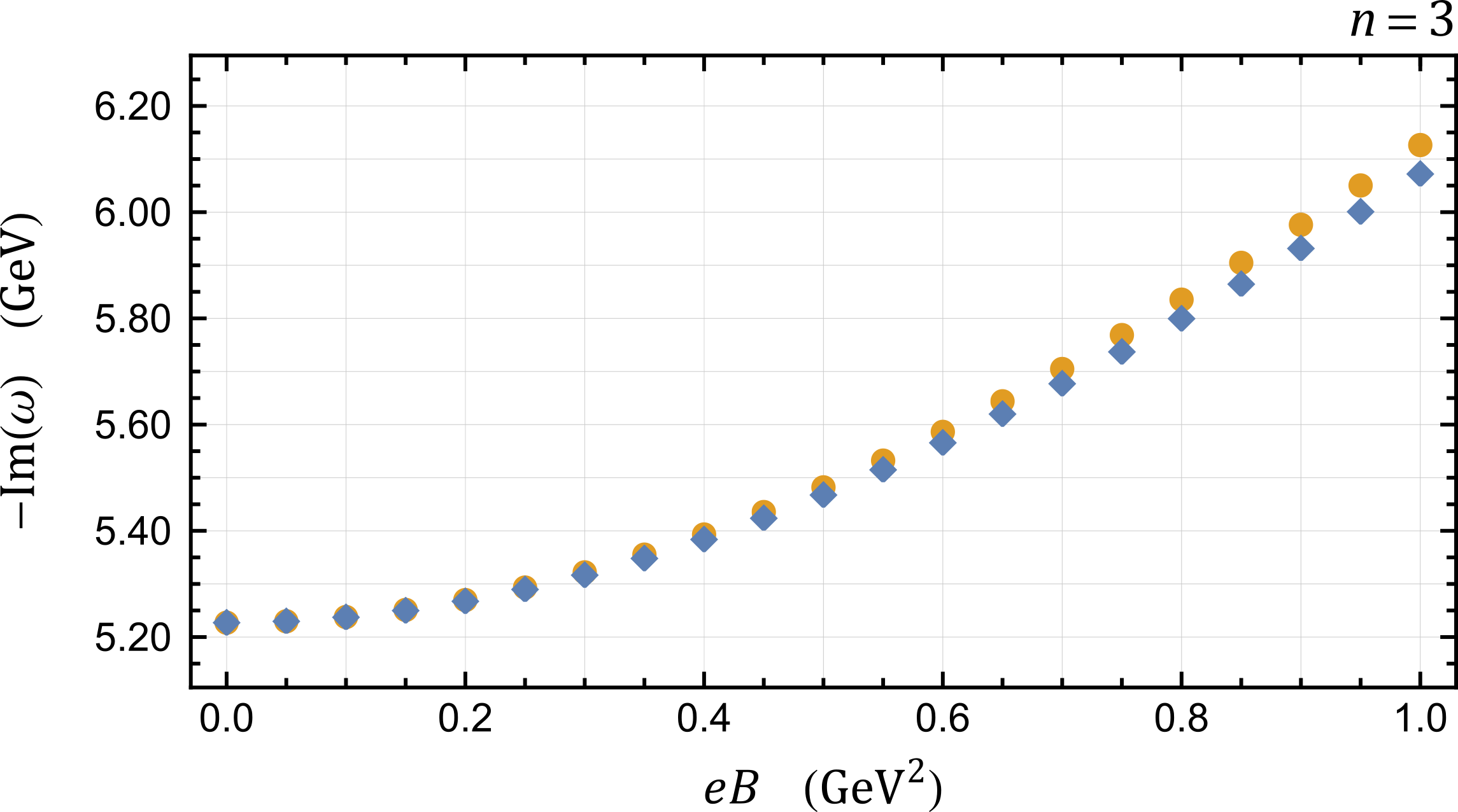}
    \\[8pt]
    \includegraphics[scale=.41]{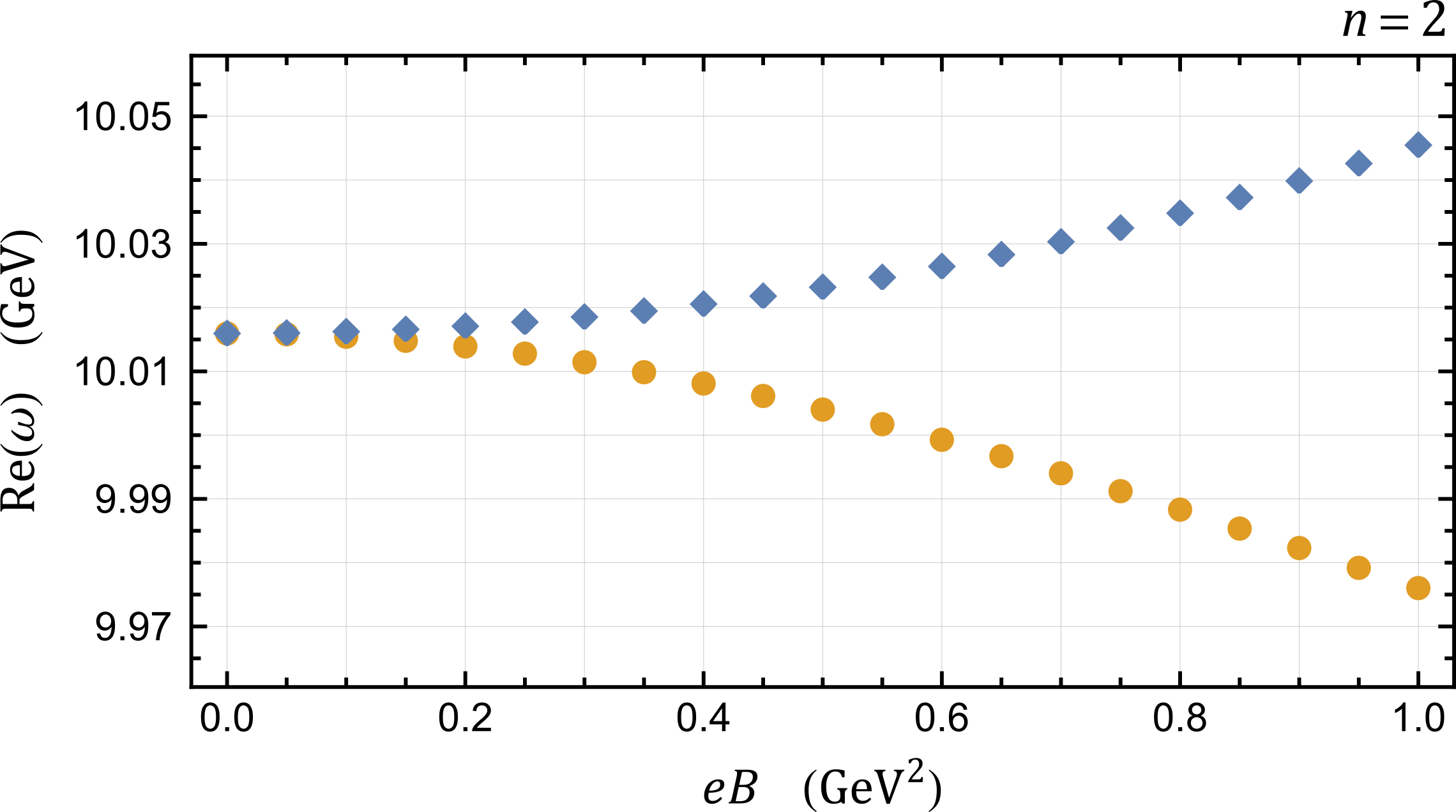} \quad
    \includegraphics[scale=.41]{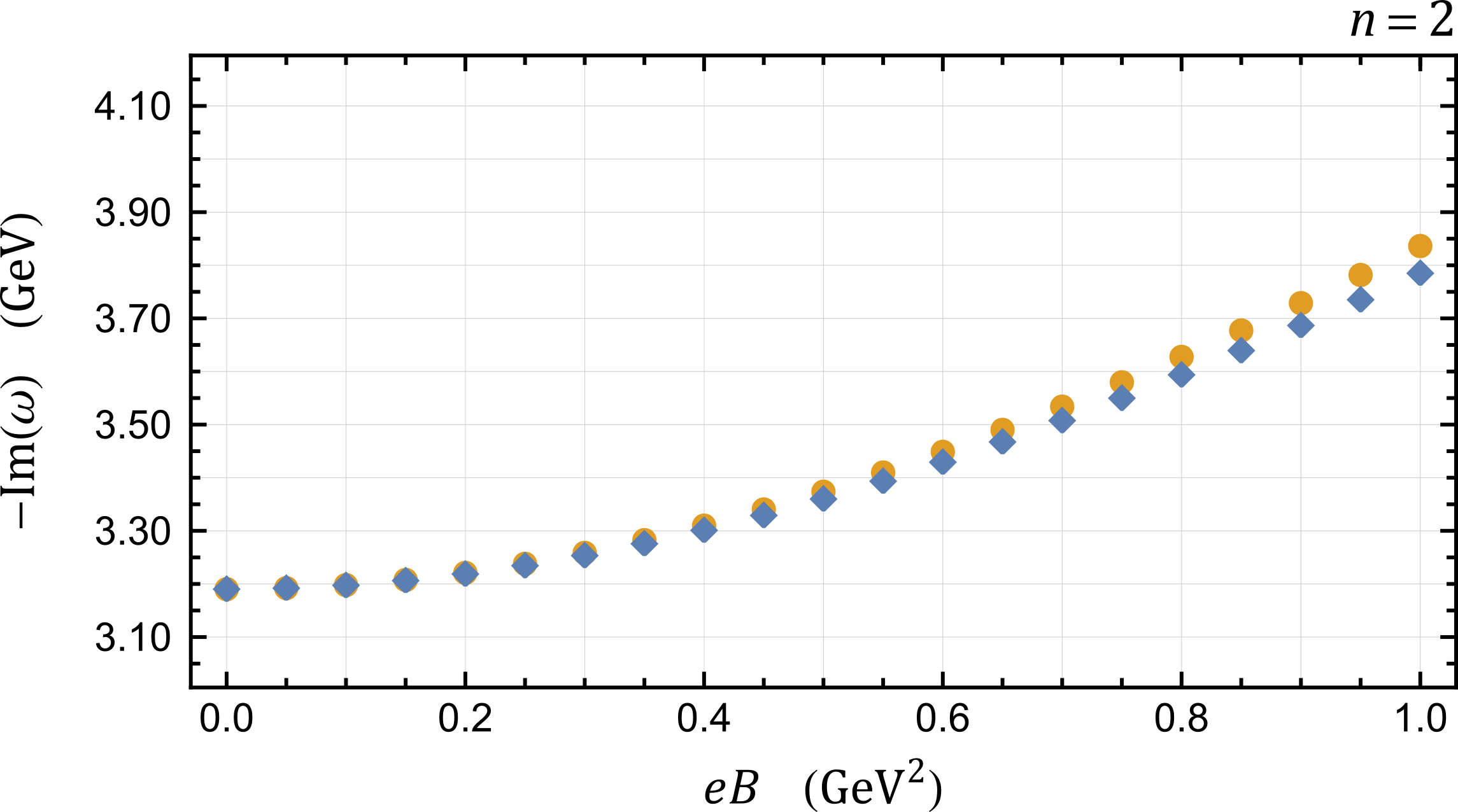}
    \\[8pt]
    \includegraphics[scale=.41]{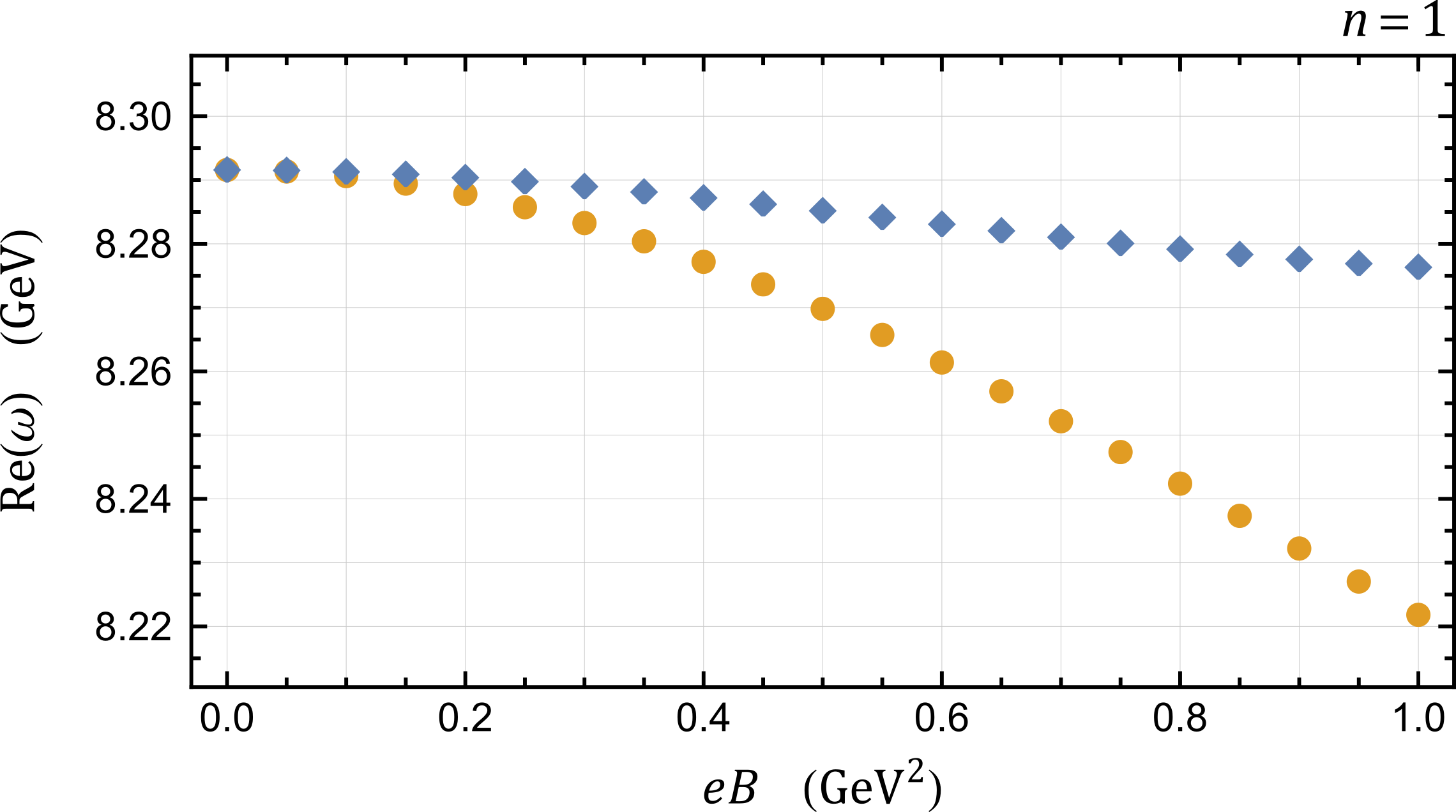} \quad
    \includegraphics[scale=.41]{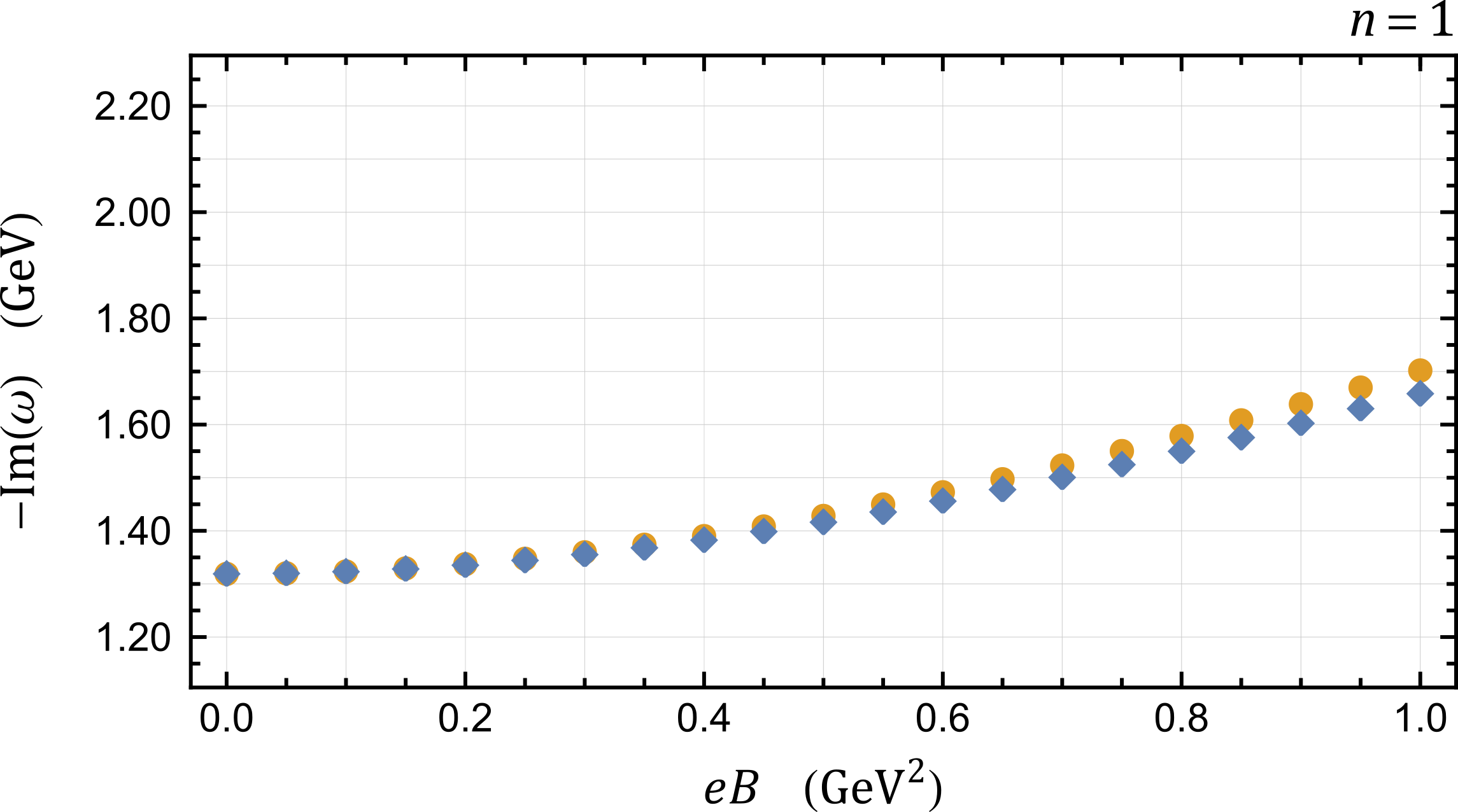}
    \\[8pt]
    \includegraphics[scale=.41]{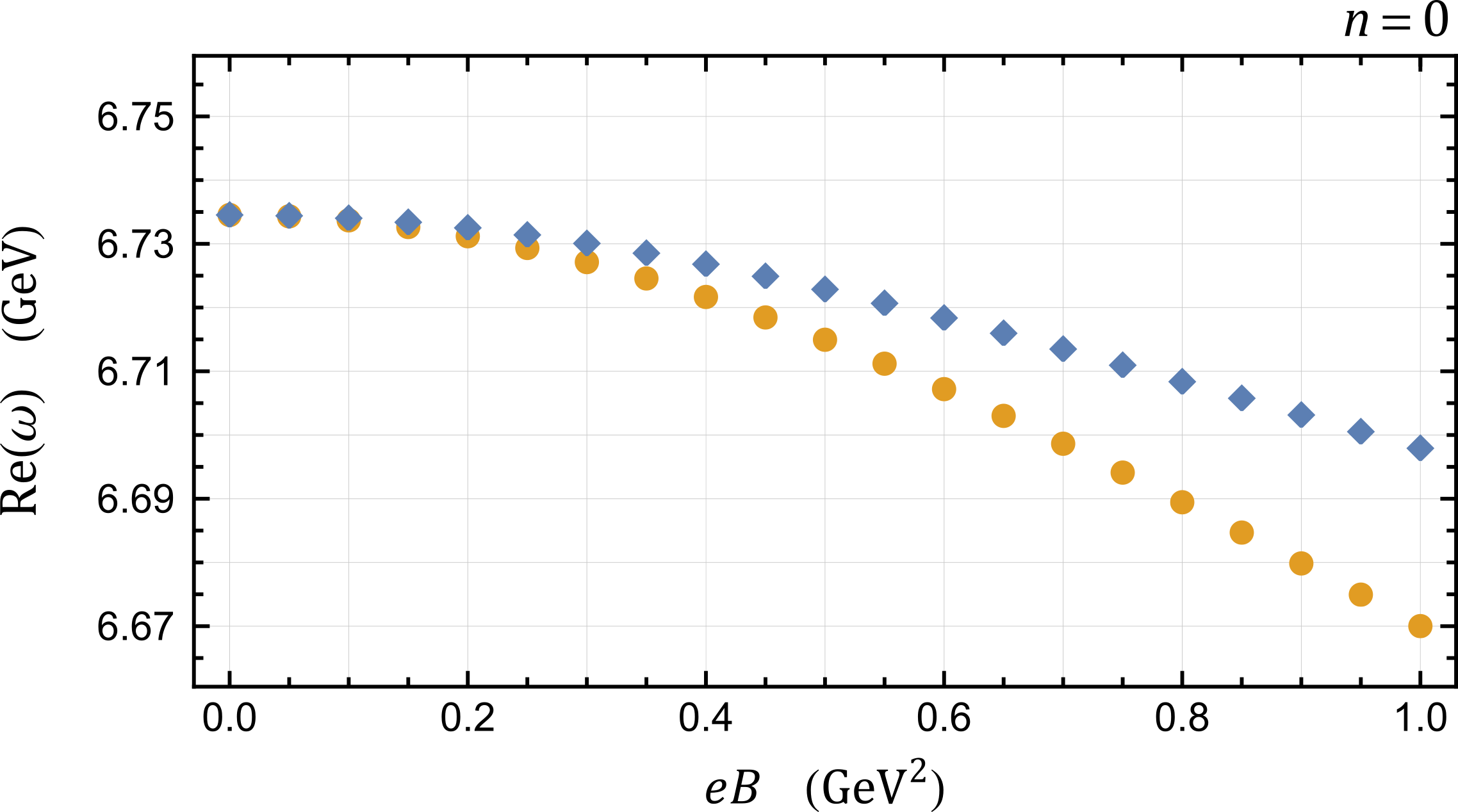} \quad
    \includegraphics[scale=.41]{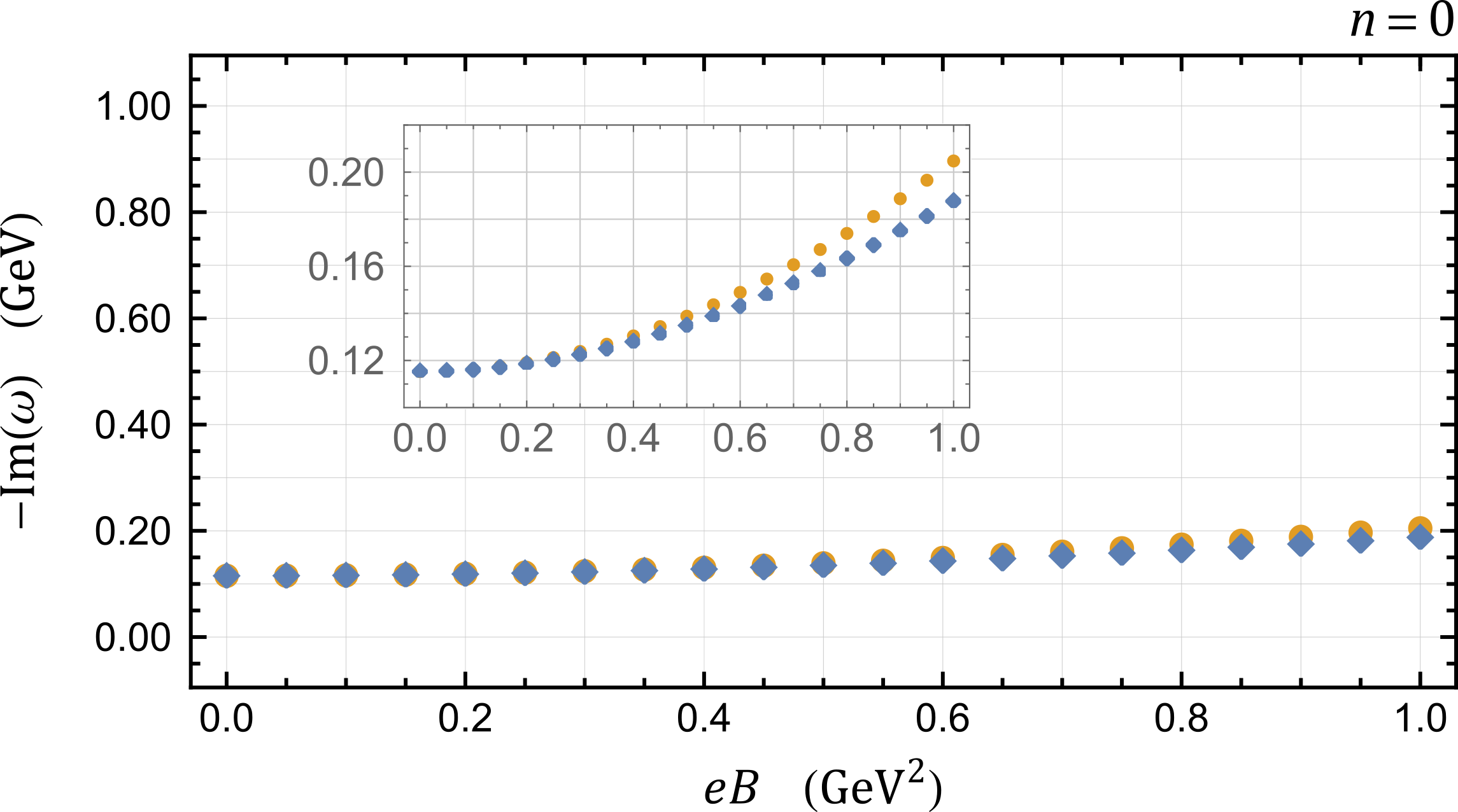}
    \\[8pt]
    \raisebox{2pt}{\includegraphics[scale=.41]{figs/MarkDiamond.png}} \quad Transverse Polarization\\[3pt]
    \raisebox{2pt}{\includegraphics[scale=.41]{figs/MarkDisk.png}}    \quad Longitudinal Polarization
    \caption{Quasinormal frequencies for transverse and longitudinal polarizations of the different excitation levels as function of the magnetic field $e B$ and for $T = \SI{300}{\mega\eV}$. All the plots of the real part of $ω$ have the same scale in order to make it easier the comparison of the variations. The same holds for the plots of the imaginary part.}
    \label{fig: QNMs2}
\end{figure}

The results of quasinormal frequencies for transverse and longitudinal polarizations as function of the magnetic field $B$ and for temperature fixed at $T = \SI{300}{\mega\eV}$ are shown in figures \ref{fig: QNMs1} and \ref{fig: QNMs2}, respectively.

The real part of the quasinormal frequency is interpreted as the thermal mass of the quasiparticle and the imaginary part is related to the degree of dissociation. The larger the absolute value of the imaginary part, the stronger the dissociation.

% ----------------------------------------------------------
\vspace{.75\baselineskip}
\section{Energy density}
\label{sec: energy density}

The energy density $ρ(z)$ of our system is the $T_{00}$ component of the energy-momentum tensor which, as in general relativity, can be calculated from
\begin{gather}
    T_{mn}(z) = \dfrac{2}{√{-g}}\COL{ \dfrac{∂(√{-g}\,\calL)}{∂g^{mn}} - ∂_p\dfrac{∂(√{-g}\,\calL)}{∂(∂_pg^{mn})} }.
\end{gather}
For transverse polarization fields, the energy density becomes
\begin{alignat}{2}
    ρ(z) &= -\dfrac{\mathrm{e}^{2\,\Im(ω)t}}{2 g_5^2 R^2}\dfrac{z^2 \mathrm{e}^{-ϕ(z)}}{d(z)}
            \PAR[3]{|ω v(z)|^2 + f(z)^2|v'(z)|^2}
    \label{eq: ρ T}
\end{alignat}
and for longitudinal polarization fields, it becomes
\begin{alignat}{2}
    ρ(z) &= -\dfrac{\mathrm{e}^{2\,\Im(ω)t}}{2 g_5^2 R^2}\dfrac{z^2 \mathrm{e}^{-ϕ(z)}}{h(z)}
            \PAR[3]{|ω v(z)|^2 + f(z)^2|v'(z)|^2}.
    \label{eq: ρ L}
\end{alignat}

The solutions of the transverse and longitudinal equations of motion can be written in the form
\begin{gather}
    v(z) = \PAR[3]{ 1 - \dfrac{z}{z_h} }^{\!\!-îω/4πT}
    \sum_{n=0}^{∞} a_n \PAR{1 - \dfrac{z}{z_h}}^{\!\!n},
    \label{eq: v series}
\end{gather}
which is the limit for $p \rightarrow ∞$ of equation \eqref{eq: vhor}. Again, $a_0 = 1$ and the other $a_n$ coefficients are determined recursively from the equation of motion and the numerical evaluation is actually performed with a finite number of terms that is determined by checking the convergence of the procedure.

It is possible to write $ρ(z)$ in the form
\begin{gather}
\label{ED}
    ρ(z) = \PAR[3]{ 1 - \dfrac{z}{z_h} }^{\!\Im(ω)/2πT}
           \sum_{n=0}^{∞} b_n \PAR[3]{ 1 - \dfrac{z}{z_h} }^{\!\!n},
\end{gather}
where the series is obtained by substituting the field given by \eqref{eq: v series}, taking the Taylor series of the factors that accompany $|v|^2$ and $|v'|^2$, multiplying the series accordingly to \eqref{eq: ρ T} or \eqref{eq: ρ L} and ordering the terms.

If we now try to compute the integral $∫_{0}^{z_h} ρ(z) ｄz$ we will have a problem: due to the singular\footnote{Remember that $\Im(ω)$ is a negative value.} factor $\PAR{ 1 - z/z_h }^{\Im(ω)/2πT}$, the integral diverges for any value of $\Im(ω)$ whose absolute value is greater than $2πT$. In fact, $\PAR{ 1 - z/z_h }^{\Im(ω)/2πT}$ behaves near $z = z_h$ as $x^{\Im(ω)/2πT}$ behaves near $x = 0$, and $∫_0^b x^{-a} ｄx$ diverges for $a ≥ 1$ and for any $b$. The energy density is, therefore, a non-normalizable function if $|\Im(ω)| ≥ 2πT$ and it happens in the second excited states and above.\footnote{See figure \ref{fig: QNMs2}. There we can find that for $n = 2$ and $n = 3$, $|\Im(ω)|$ is always bigger than $2π × \SI{300}{\mega\eV} ≃ \SI{1.88}{\giga\eV}$} One cannot use a non-normalizable function to define the modal fraction. So, we extract the factor $\PAR{ 1 - z/z_h }^{\Im(ω)/2πT}$, which is responsible for the divergence, keeping only the regular factor. This is achieved by simply introducing
\begin{gather}
    R(z) = \PAR{ 1 - \dfrac{z}{z_h} }^{\!\!-\Im(ω)/2πT} ρ(z) = \sum_{n=0}^{∞} b_n \PAR[3]{ 1 - \dfrac{z}{z_h} }^{\!\!n},
    \label{eq: R(z)}
\end{gather}
whose Fourier transform is
\begin{align}
    \tilde{R}(k) &= ∫_{0}^{z_h^{}} R(z)\, \mathrm{e}^{-î k z}\, ｄz \,.
    \label{eq: fourier transform of R(z)}
\end{align}
Then we define the modal fraction as
\begin{gather}
    ϵ(k) = \dfrac{ |\tilde{R}(k)|^2 }{\displaystyle \Max(|\tilde{R}(k)|^2) \rule{0ex}{2.4ex}}.
    \label{eq: modal fraction R}
\end{gather}
The maximum value of $|\tilde{R}(k)|^2$ is at $k = 0$.

An important detail of the numerical calculation that is worth noting is that, due to the highly oscillatory behavior of the factor $(1 - z/z_h )^{-îω/4πT}$ near $z = z_h$, we do not evaluate the field $v$ in the region $z_0 < z < z_h$,. However, we can evaluate $R(z)$ directly in this region by using the series expansion in \eqref{eq: R(z)} and the fact that the coefficients $b_n$ of the energy density can be written in terms of the coefficients $a_n$ of the field.

% ----------------------------------------------------------
\vspace{.75\baselineskip}
\section{DCE from regularized energy density}
\label{sec: CE}

\begin{figure}[htb!]
    \centering
    \includegraphics[scale=.45]{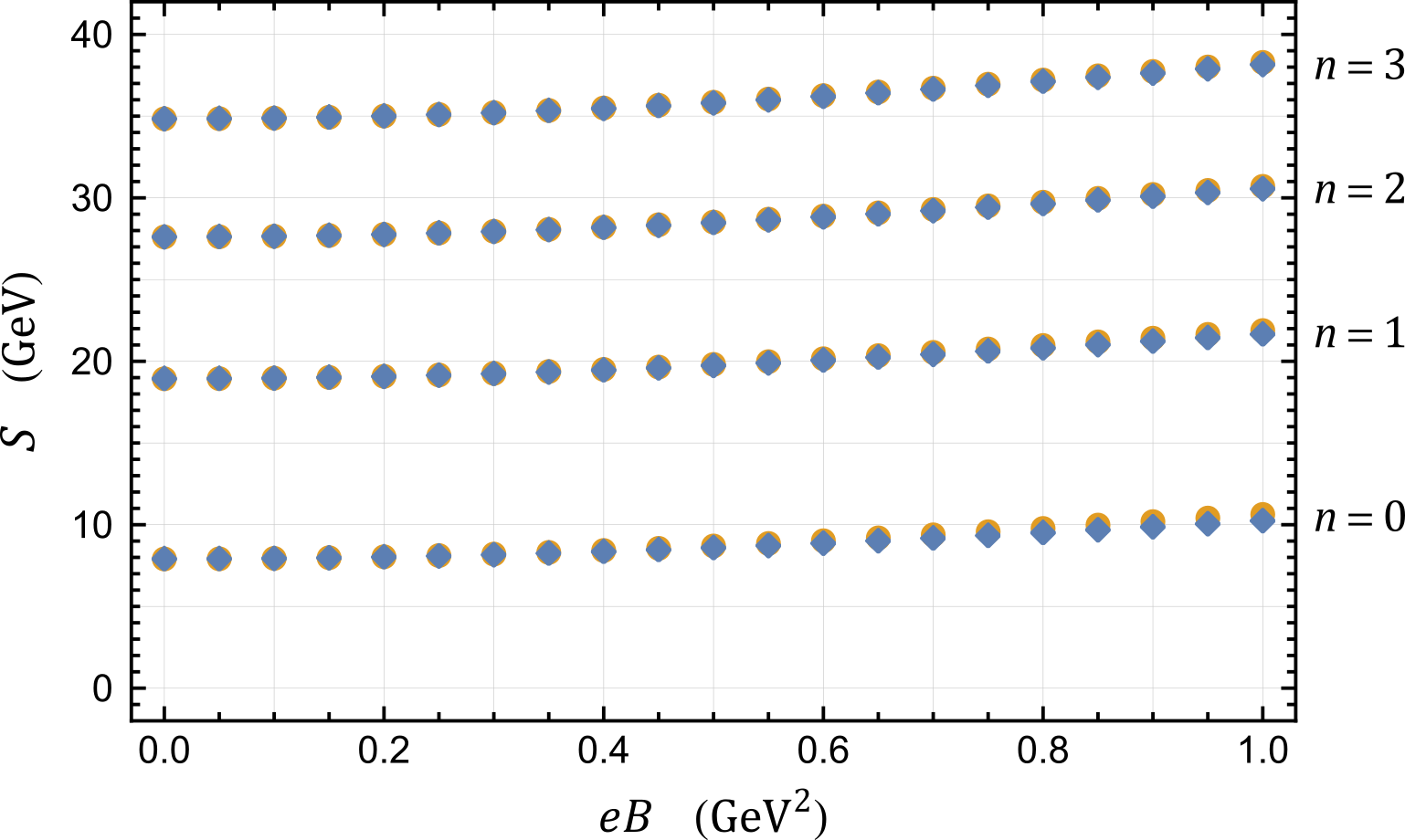}
    \\[8pt]
	\raisebox{2pt}{\includegraphics[scale=.36]{figs/MarkDiamond.png}} \quad Transverse Polarization\\[3pt]
    \raisebox{2pt}{\includegraphics[scale=.36]{figs/MarkDisk.png}}    \quad Longitudinal Polarization
    \caption{Differential configuration entropy for transverse and longitudinal polarizations of the different excitation levels as a function of the magnetic field $eB$ at $T = \SI{300}{\mega\eV}$.}
    \label{fig: CEs1}
\end{figure}

Using the numerical solutions described in the last section, one obtains the modal fraction $ϵ(k)$ from equation \eqref{eq: modal fraction R} and then finds the DCE
\begin{gather}
    S = -∫_{-∞}^{+∞} ϵ(k) \ln ϵ(k)\, ｄk.
\end{gather}
It is important to remark that the quasinormal mode solutions, and consequently the DCE, depend on $T$ and $B$. We fixed the temperature at $T = \SI{300}{\mega\eV}$, a value of $T$ in which the plasma is present but the bottomonium states are only partially dissociated. The result obtained for the DCE as function of $eB$ is shown in figure \ref{fig: CEs1}, where we plot the four lowest radial modes $n = 0,1,2,3$ together in order to show the variation of the DCE with the excitation level. From this figure it is observed that the DCE increases with the excitation level. A result that is consistent with the fact that the higher order states are more unstable against dissociation in the thermal medium.

\begin{figure}[htb!]
    \centering
    \vspace{2\baselineskip}
    \includegraphics[scale=.43]{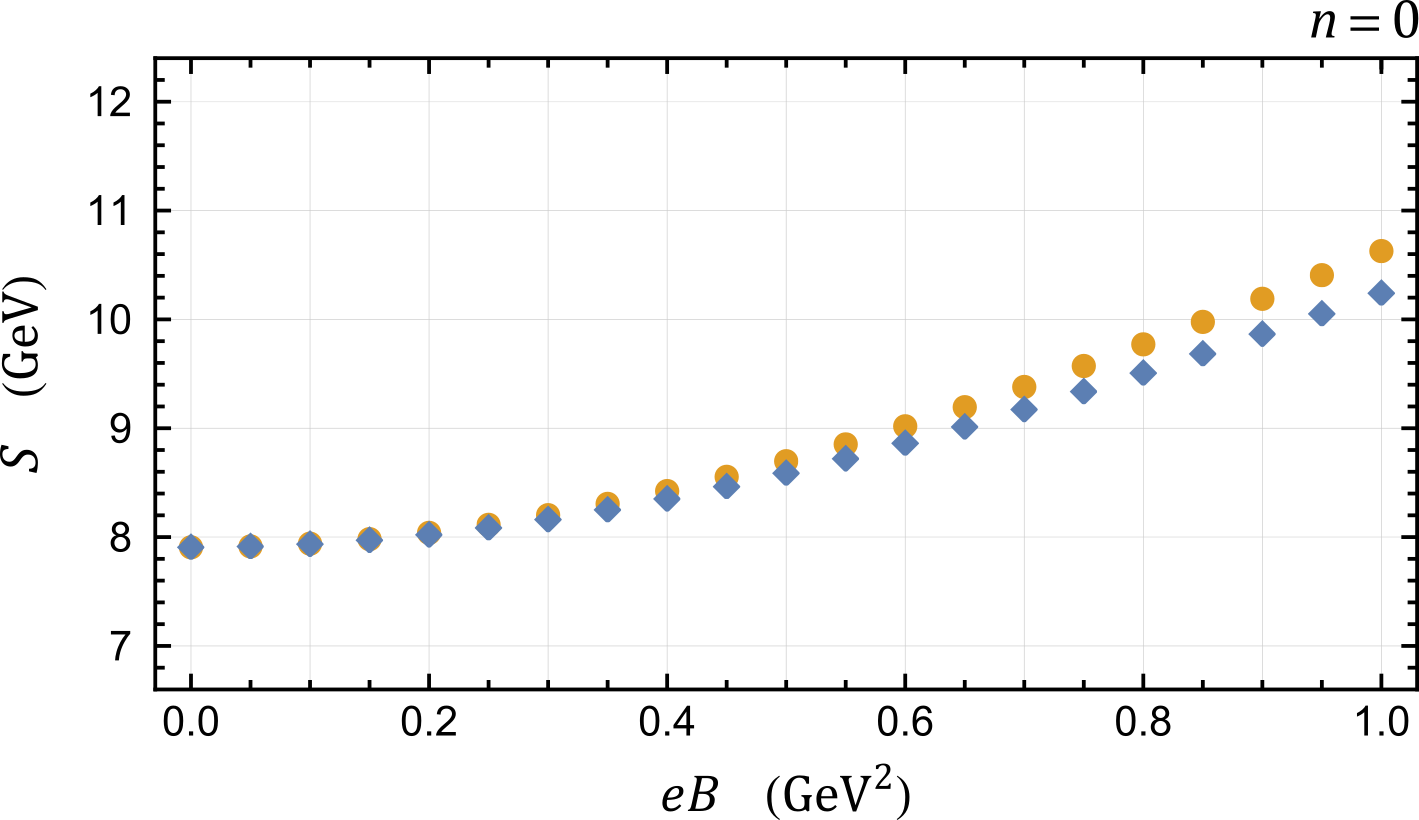} \quad
    \includegraphics[scale=.43]{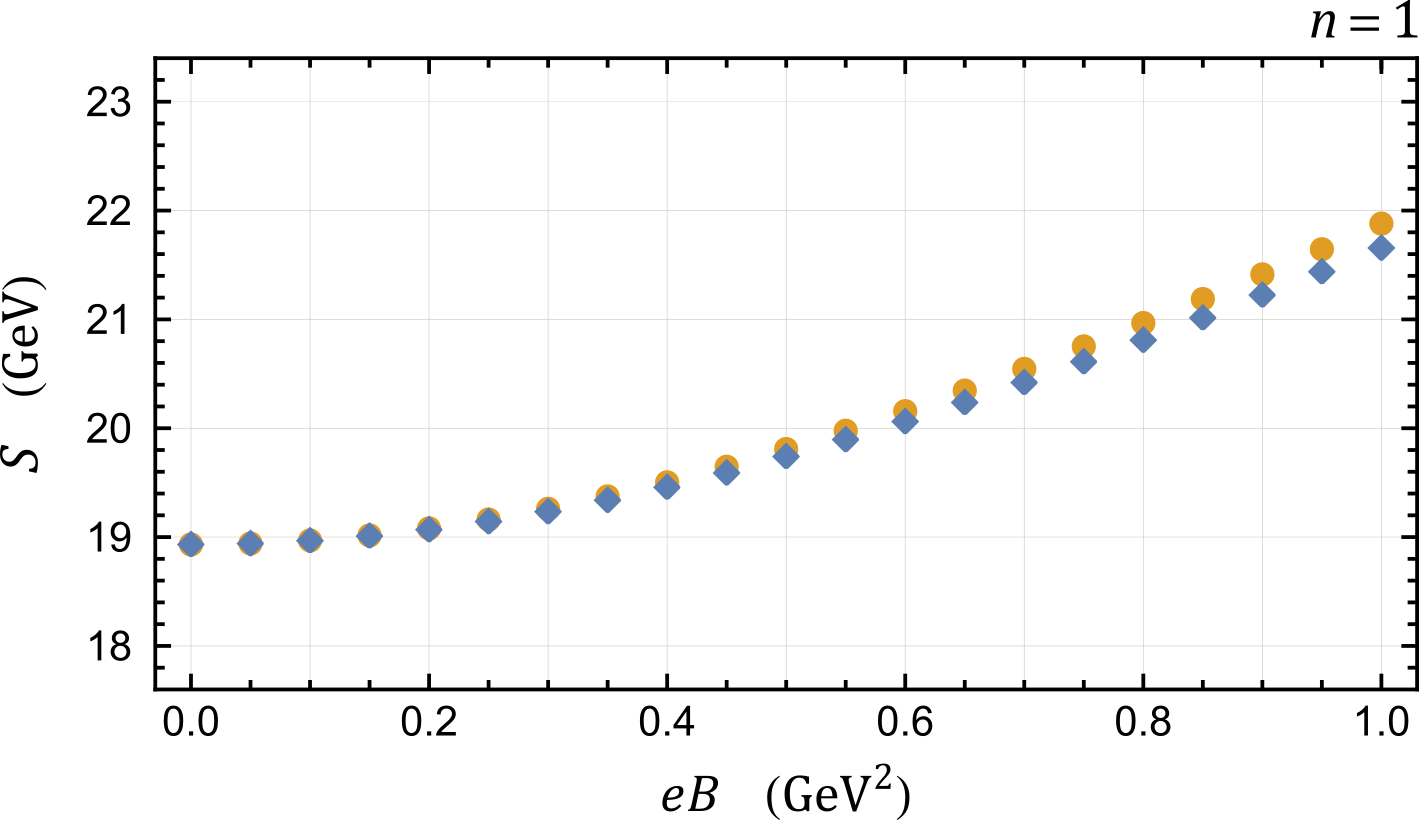}
    \\[8pt]
    \includegraphics[scale=.43]{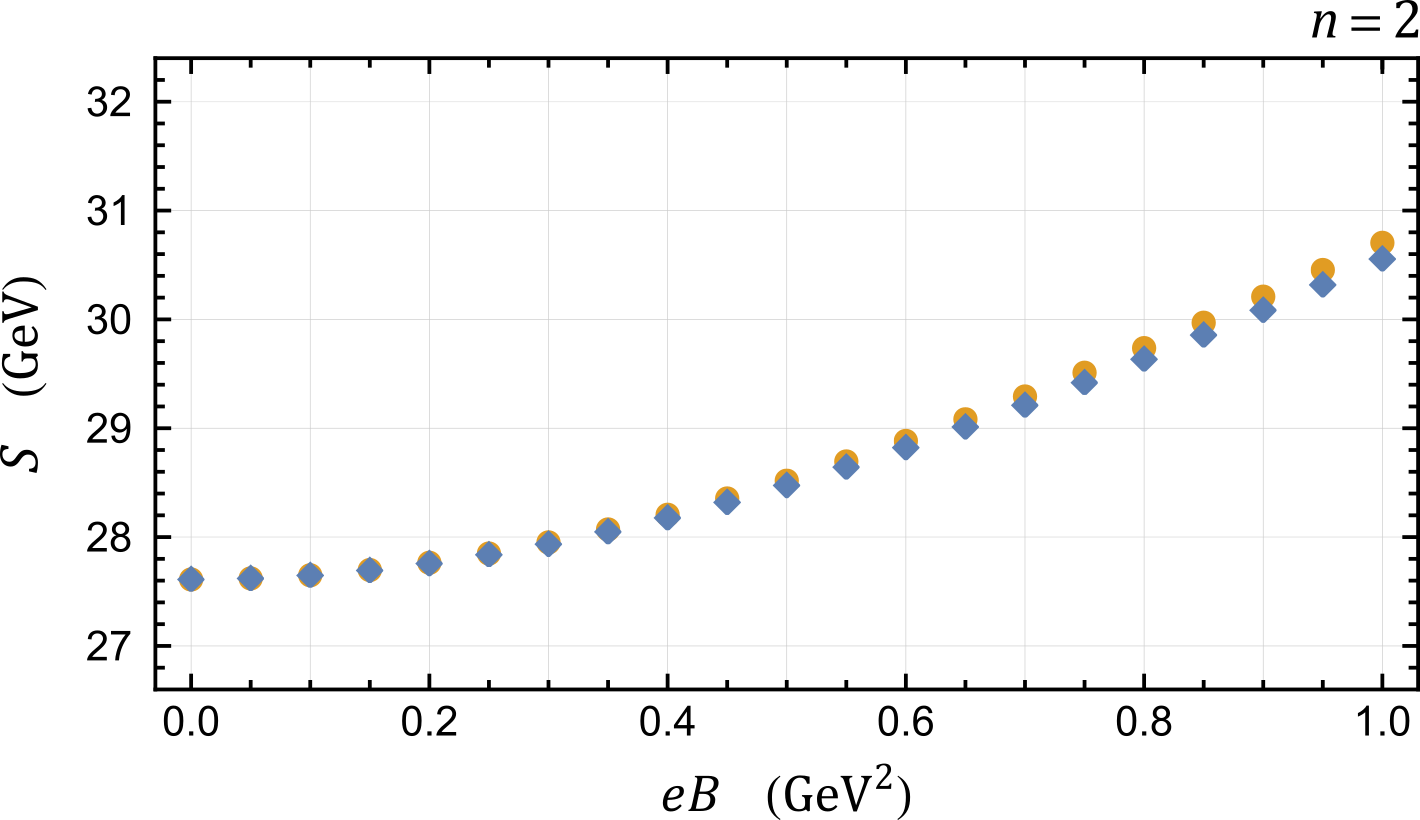} \quad
    \includegraphics[scale=.43]{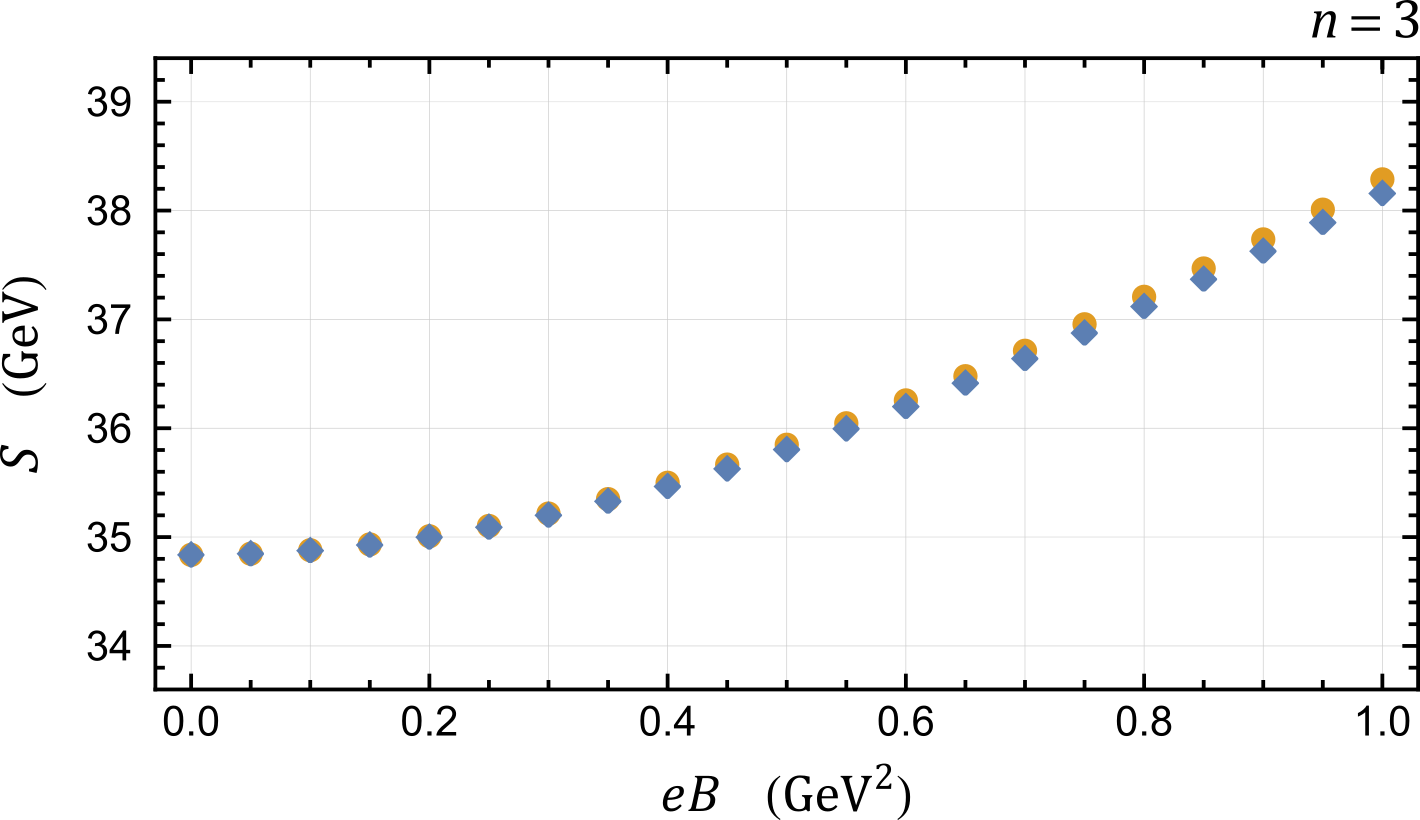}
    \\[8pt]
	\raisebox{2pt}{\includegraphics[scale=.36]{figs/MarkDiamond.png}} \quad Transverse Polarization\\[3pt]
    \raisebox{2pt}{\includegraphics[scale=.36]{figs/MarkDisk.png}}    \quad Longitudinal Polarization
    \caption{Differential configuration entropy for transverse and longitudinal polarizations of the $n = 0,1,2,3$ excitation levels as a function of the field $eB$ for $T = \SI{300}{\mega\eV}$. All the plots of the DCE have the same scale in order to make it easier the comparison of the variations.}
    \label{fig: CEs2}
\end{figure}

Then, in the four panels of figure \ref{fig: CEs2}, we show the DCE for each of the four lowest radial modes, with an expanded scale in order to make it possible to notice the variation with the magnetic field and the split between longitudinal and transverse modes. From this figure one notices that the DCE increases with the magnetic field, indicating an enhancement of the dissociation in the medium. This effect is slightly larger for the longitudinal polarization case.

Remarkably, the dependence of the DCE on the field $eB$ is found to be of a simple form. The results obtained can be fitted by a polynomial of degree 2 in the magnetic field
\begin{equation}
\label{fit}
S = c_0 + c_1 (eB) + c_2 (eB)^2 \,,
\end{equation}
with values of the adjusted coefficient of determination, $R^2_{\adj}$, very close to one, indicating a very nice fit. We show the results of the fit in tables \ref{tab: CEsT fit} and \ref{tab: CEsL fit}.

\begin{table}[htb!]
\centering
%\vspace{\baselineskip}
\begin{tabular}{
    | >{ \,$}c<{$\, }                   % n   column
    | >{ \,$}c<{$\, }                   % c_0 column
    | >{ \,$}c<{$\, }                   % c_1 column
    | >{ \,$}c<{$\, }                   % c_2 column
    | >{ \,$}c<{$\, }                   % R^2 column
    | }
    \hline
      n                                 % n   column
    & c_0 \ (\si{\giga\eV})             % c_0 column
    & c_1 \ (\si{\giga\eV^{-1}})        % c_1 column
    & c_2 \ (\si{\giga\eV^{-3}})        % c_2 column
    & R^2_{\adj} \rule[-7pt]{0pt}{22pt} % R^2 column
    \\ \hline
    1 & 07.87 \pm 0.01 & 0.46 \pm 0.06 & 1.95 \pm 0.06 & 0.9999942  \rule[-1pt]{0pt}{12pt} \\
    2 & 18.89 \pm 0.02 & 0.59 \pm 0.07 & 2.23 \pm 0.07 & 0.9999981  \rule[-1pt]{0pt}{12pt} \\
    3 & 27.56 \pm 0.02 & 0.61 \pm 0.08 & 2.44 \pm 0.07 & 0.9999991  \rule[-1pt]{0pt}{12pt} \\
    4 & 34.78 \pm 0.02 & 0.66 \pm 0.09 & 2.77 \pm 0.08 & 0.9999992  \rule[-1pt]{0pt}{12pt} \\ \hline
\end{tabular}
\caption{Coefficients of the fit of the DCE for transverse polarization of the different excitation levels with the form given in equation (\ref{fit}) for temperature fixed at $T = \SI{300}{\mega\eV}$.}
\label{tab: CEsT fit}
\end{table}

\begin{table}[htb!]
\centering
\vspace{.5\baselineskip}
\begin{tabular}{
    | >{ \,$}c<{$\, }                   % n   column
    | >{ \,$}c<{$\, }                   % c_0 column
    | >{ \,$}c<{$\, }                   % c_1 column
    | >{ \,$}c<{$\, }                   % c_2 column
    | >{ \,$}c<{$\, }                   % R^2 column
    | }
    \hline
      n                                 % n   column
    & c_0 \ (\si{\giga\eV})             % c_0 column
    & c_1 \ (\si{\giga\eV^{-1}})        % c_1 column
    & c_2 \ (\si{\giga\eV^{-3}})        % c_2 column
    & R^2_{\adj} \rule[-7pt]{0pt}{22pt} % R^2 column
    \\ \hline
    1 & 07.86 \pm 0.01 & 0.53 \pm 0.07 & 2.28 \pm 0.07 & 0.9999925 \\
    2 & 18.88 \pm 0.02 & 0.65 \pm 0.08 & 2.40 \pm 0.08 & 0.9999978 \\
    3 & 27.56 \pm 0.02 & 0.64 \pm 0.08 & 2.56 \pm 0.08 & 0.9999990 \\
    4 & 34.78 \pm 0.02 & 0.71 \pm 0.09 & 2.86 \pm 0.09 & 0.9999992 \\ \hline
\end{tabular}
\caption{Coefficients of the fit of the DCE for longitudinal polarization of the different excitation levels with the form given in equation (\ref{fit}) for temperature fixed at $T = \SI{300}{\mega\eV}$.}
\label{tab: CEsL fit}
\end{table}

\begin{figure}[htb!]
    \centering
    % \vspace{\baselineskip}
    \includegraphics[scale=.45]{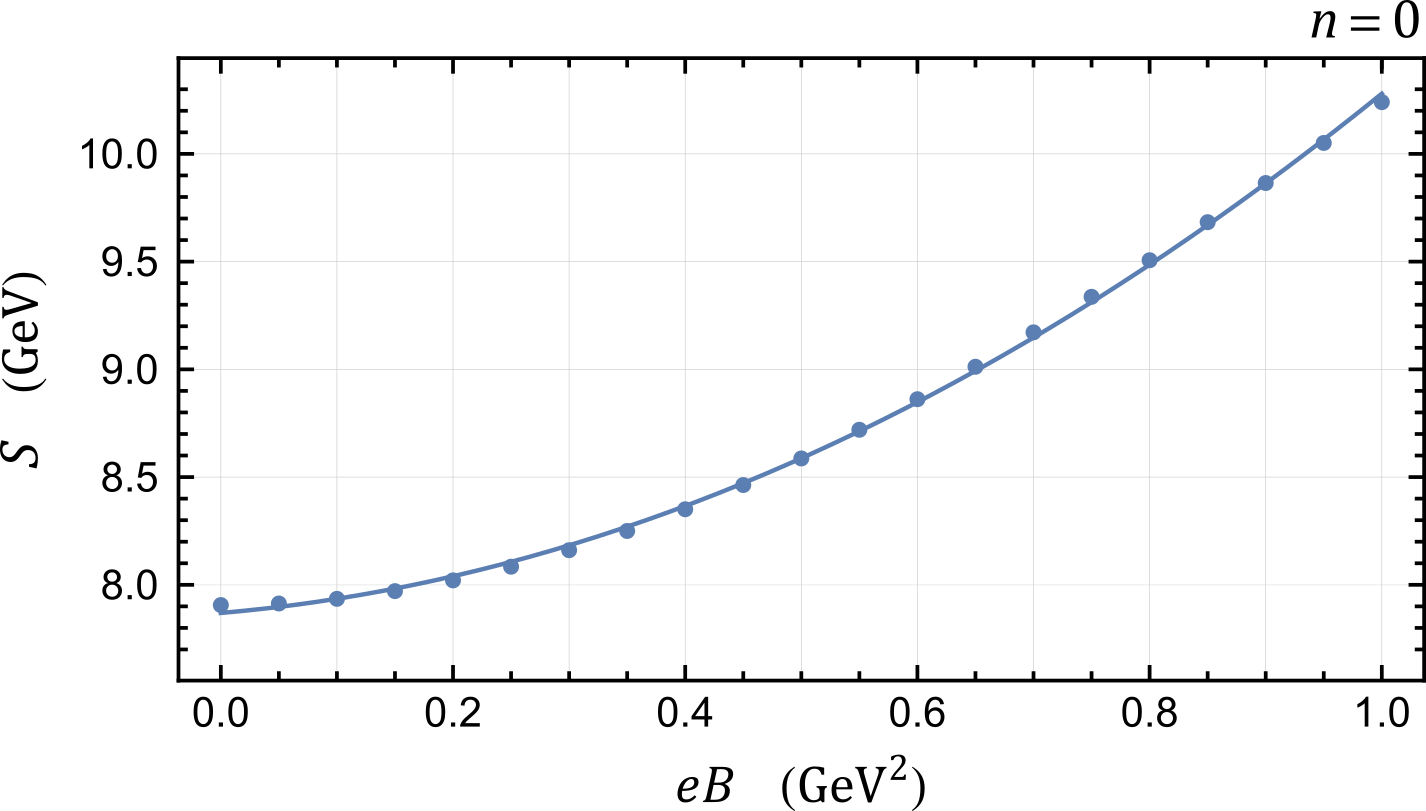}
    \caption{Quadratic fit of the differential configuration entropy of bottomonium field with transverse polarizations in the ground state as a function of the magnetic field $eB$ for $T = \SI{300}{\mega\eV}$.}
    \label{fig: CE Fit}
\end{figure}

As an illustration of the quadratic fit, we show in Figure \ref{fig: CE Fit} the actual values of the DCE, represented by points, and a line corresponding to the fit. We have taken the case of transverse polarization and $n = 0$ as an example. The plots for the other states are similar.

% ----------------------------------------------------------
\vspace{.75\baselineskip}
\section{DCE from nonregularized energy density}
\label{sec: NRCE}

The results obtained in the previous section are consistent with the fact that heavy vector mesons become more unstable as the background magnetic field increases. The increase in the DCE obtained from the regularized energy density of equation \eqref{eq: R(z)} indeed indicates increase in instability. However, the continuous increase in the DCE does not tell us anything about an important physical aspect. Namely, heavy vector mesons not only become more unstable but rather they completely dissociate in the medium for temperatures above some value, which depends on the magnetic field. The DCE obtained in last section does not give us any information about the magnitudes of temperature and magnetic field where the meson quasistates are no more present in the medium.

The motivation that lead us to define the DCE from the regularized energy density in the previous section is that the spatial function $R$ in equation \eqref{eq: intro - R(k)}, which one uses in order to define the modal fraction, has to be square integrable and the energy density does not satisfy this requirement in general. However, since we are dealing with the calculation of a property of quasiparticles that dissociate when $T$ and (or) $eB$ increase, it is worth checking what happens if one uses the nonregularized energy density $ρ$ of equations \eqref{eq: ρ T} and \eqref{eq: ρ L}, instead of the regularized one.

\begin{figure}[htb!]
    \centering
    \includegraphics[scale=.45]{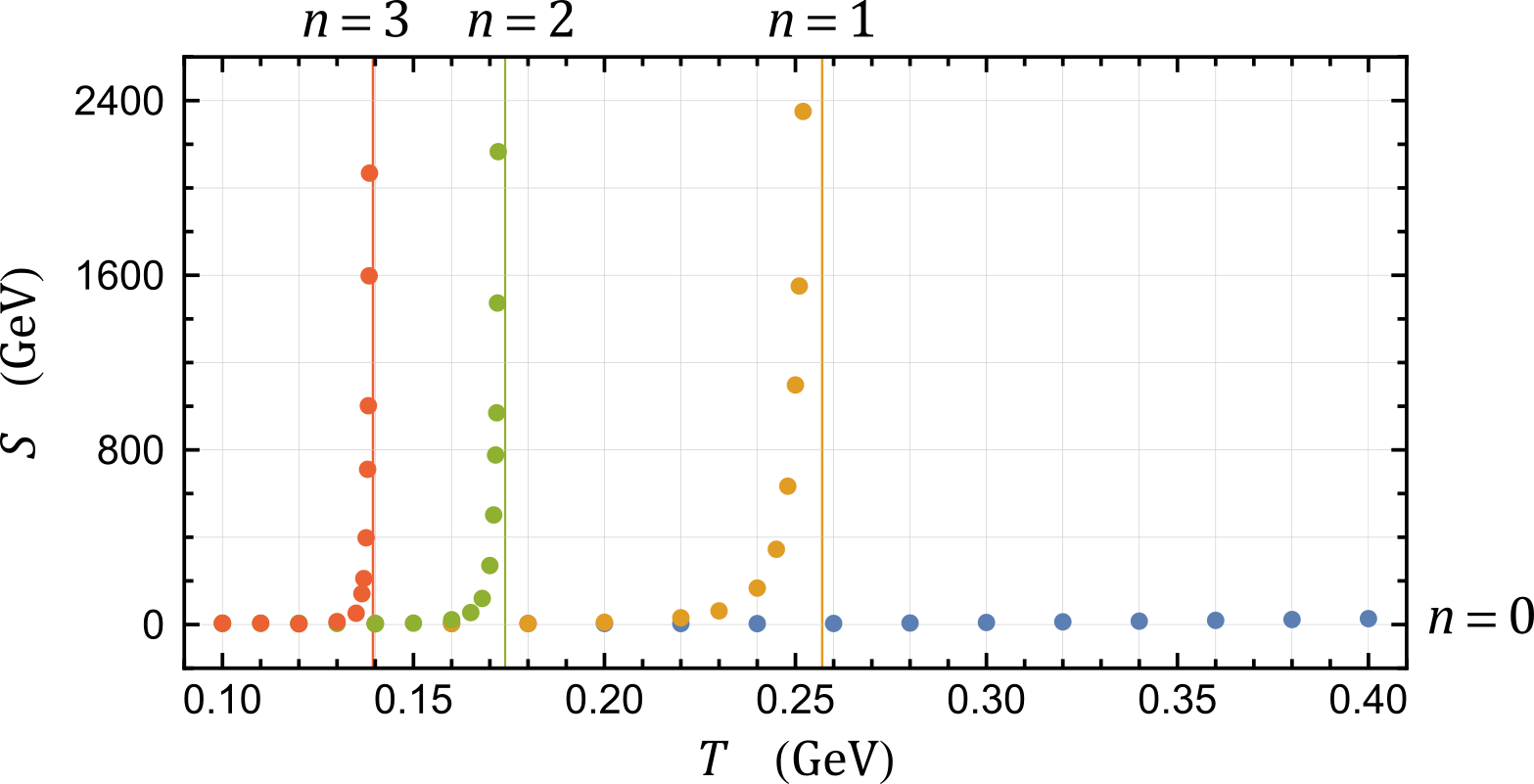}
    \caption{DCE for the $ n = 0,1,2,3 $ modes, obtained from the nonregularized energy density. The vertical lines correspond to the temperatures where the DCE diverges. }
    \label{fig: NonRegDCE}
\end{figure}

For this purpose, let us us now define
\begin{align}
    \tilde{ρ}(k) &= ∫_{0}^{z_h^{}} ρ(z)\, \mathrm{e}^{-î k z}\, ｄz \,,
    \label{eq: fourier transform of rho(z)}
\end{align}
and the modal fraction as
\begin{gather}
    ϵ(k) = \dfrac{ |\tilde{ρ}(k)|^2 }{\displaystyle \Max(|\tilde{ρ}(k)|^2) \rule{0ex}{2.4ex}}.
    \label{eq: modal fraction rho}
\end{gather}
In order to investigate the effect of using this potentially singular expression for the DCE we start by considering just a medium with temperature and not with magnetic field. Calculating numerically the DCE for the modes $n = 0, 1, 2, 3$ as a function of the temperature one finds an interesting result, which we plot in figure \ref{fig: NonRegDCE}. One notes that for the mode $ n = 1$ the DCE diverges for $T > T_{1c} = \SI{257}{\mega\eV}$, while for the $ n=2$ quasistate the divergence occurs for $ T > T_{2c} = \SI{174}{\mega\eV}$ and the $ n=3 $ diverges for $ T > T_{3c} = \SI{139}{\mega\eV}$. The DCE calculated from the nonregularized energy density is not defined for higher temperatures for these modes. For the $ n = 0$ mode the same type of divergence occurs at a higher temperature, of the order of $\SI{1}{\giga\eV}$, above the range of temperatures found in the quark gluon plasma.

One can interpret this result as telling us that the total dissociation of the quasistates corresponds to infinite instability and thus to infinite DCE. So, using the nonregularized energy density one finds a clear trace of the disappearance of the quasistates when some critical temperature, that depends on the order $n$ of the mode, is reached. The absence of a definition for the DCE for higher temperatures can be interpreted as corresponding to the fact that there is no more quasiparticles in this case.

\begin{figure}[htb!]
    \centering
    \includegraphics[scale=.45]{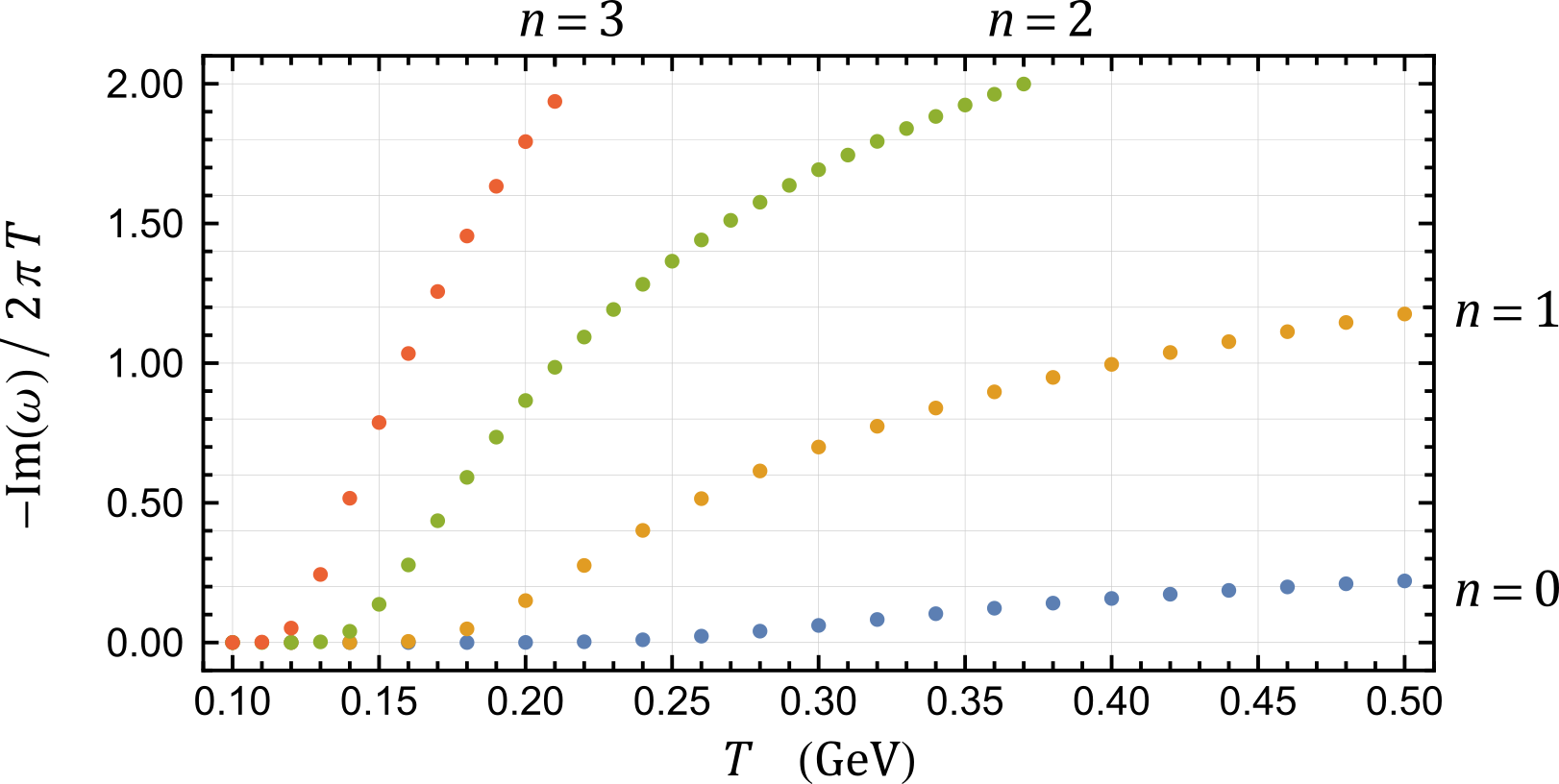}
    \caption{Values of the factor $\,-\Im(ω)/2πT$ for $ n=0,1,2,3$ as functions of the temperature.}
    \label{fig: singularfactor}
\end{figure}

The values of the critical temperatures $T_{1c}, T_{2c}$ and $T_{3c}$ correspond exactly to the values of $T$ where the corresponding $n = 1,2,3$ modes satisfy the condition:
\begin{equation}
    \label{eq: sign}
    \Im(ω)/2πT = - 0.5 \,.
\end{equation}
This happens because the energy density contains, as shown in equation \eqref{ED}, the potentially singular factor $\PAR{1 - z/z_h}^{\Im(ω)/2πT}$. So, the square of the energy density becomes singular at the horizon when the condition \eqref{eq: sign} is satisfied. Then, the energy density is no more square integrable and the DCE is not defined. We plot in figure \ref{fig: singularfactor} the variation of the factor $- \Im(ω)/2πT$ as a function of $T$ for the four modes so that one can see how the condition \eqref{eq: sign} is satisfied.

\begin{figure}[htb!]
    \centering
    \vspace{5pt}
    \includegraphics[scale=.46]{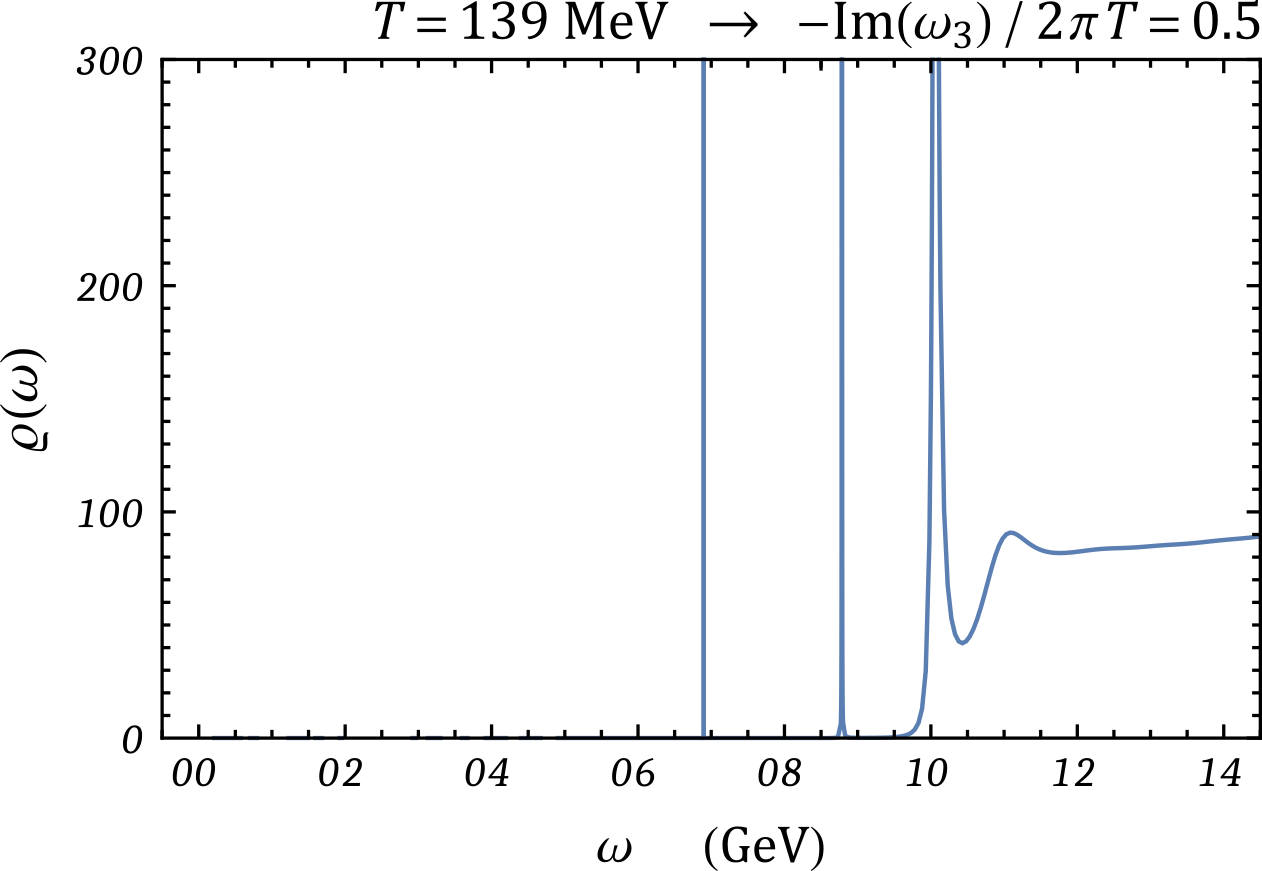} \quad
    \includegraphics[scale=.46]{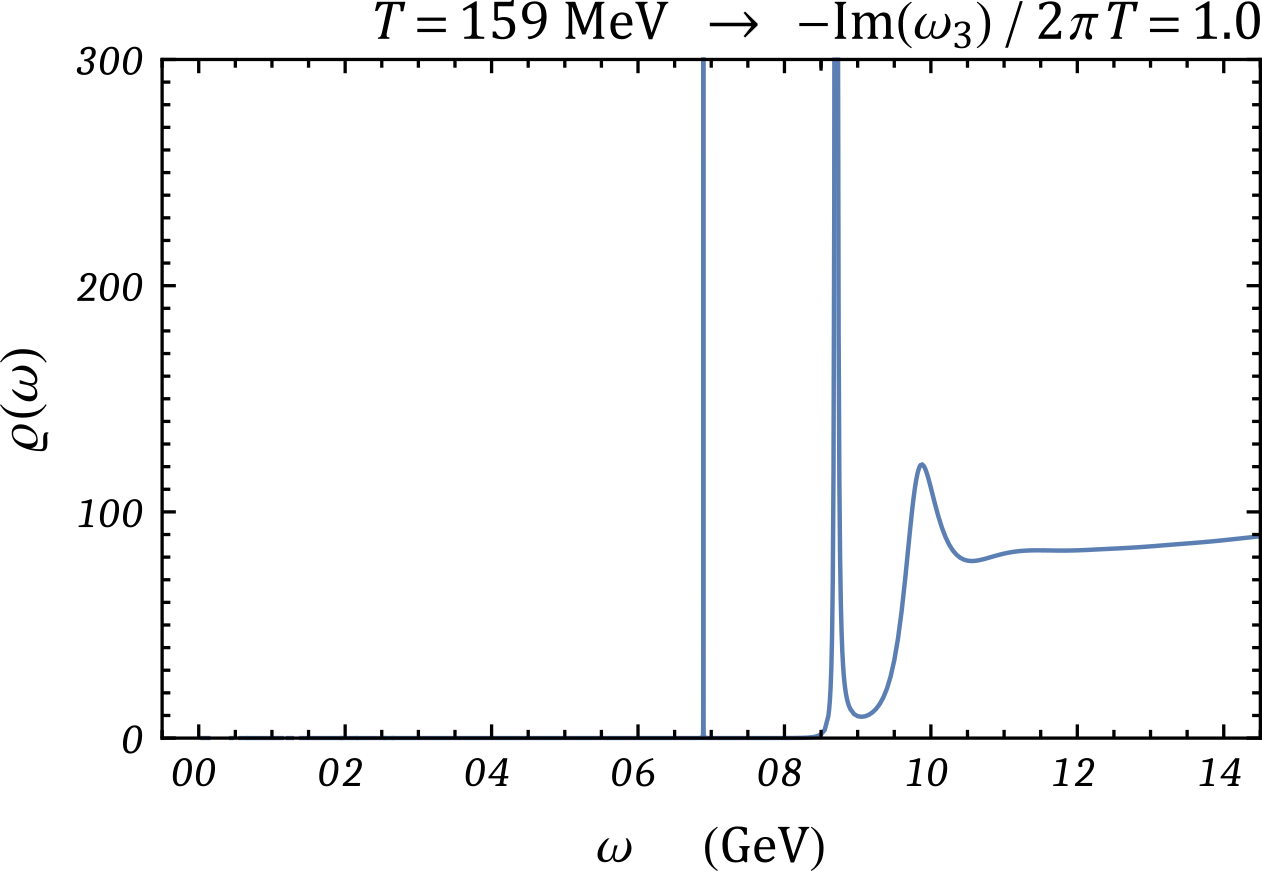}
    \\[10pt]
    \includegraphics[scale=.46]{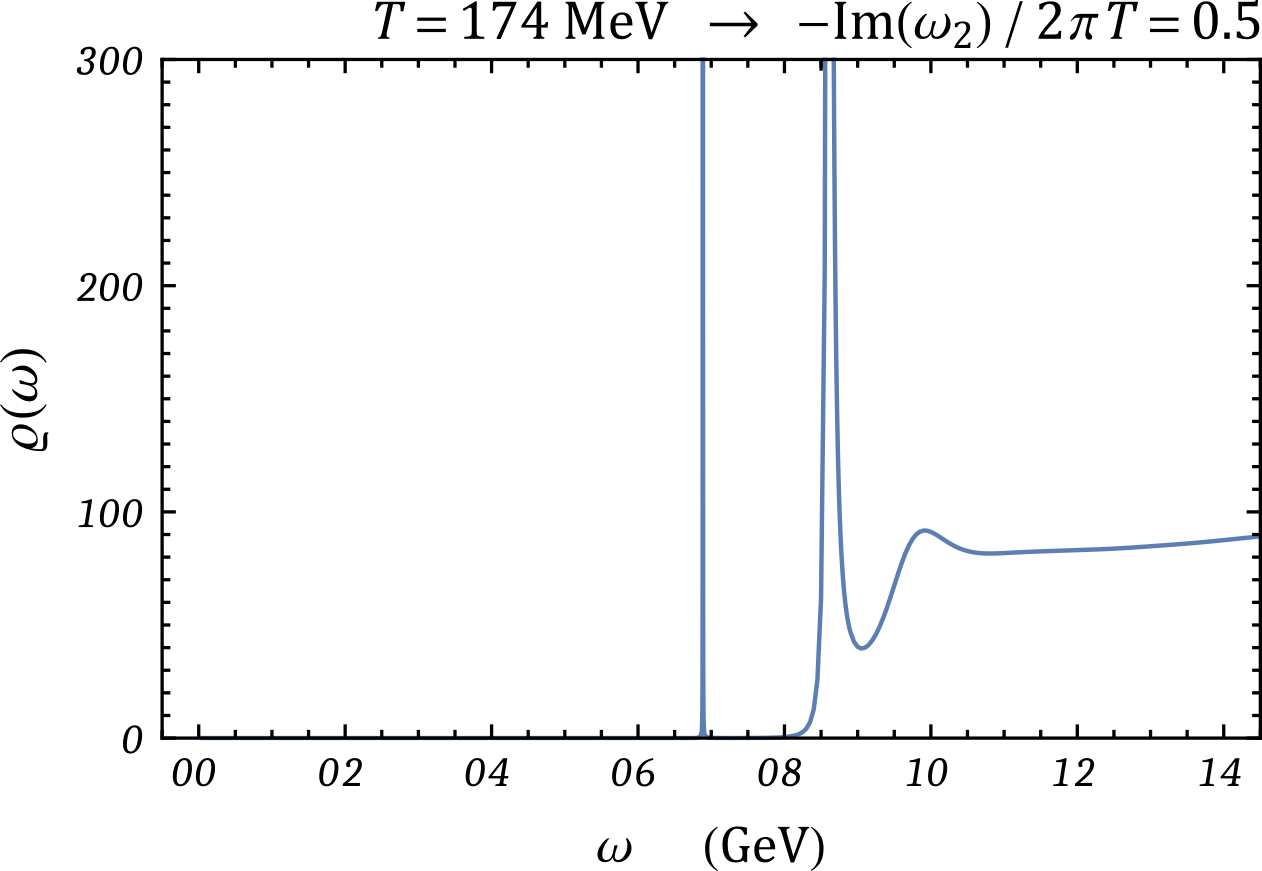} \quad
    \includegraphics[scale=.46]{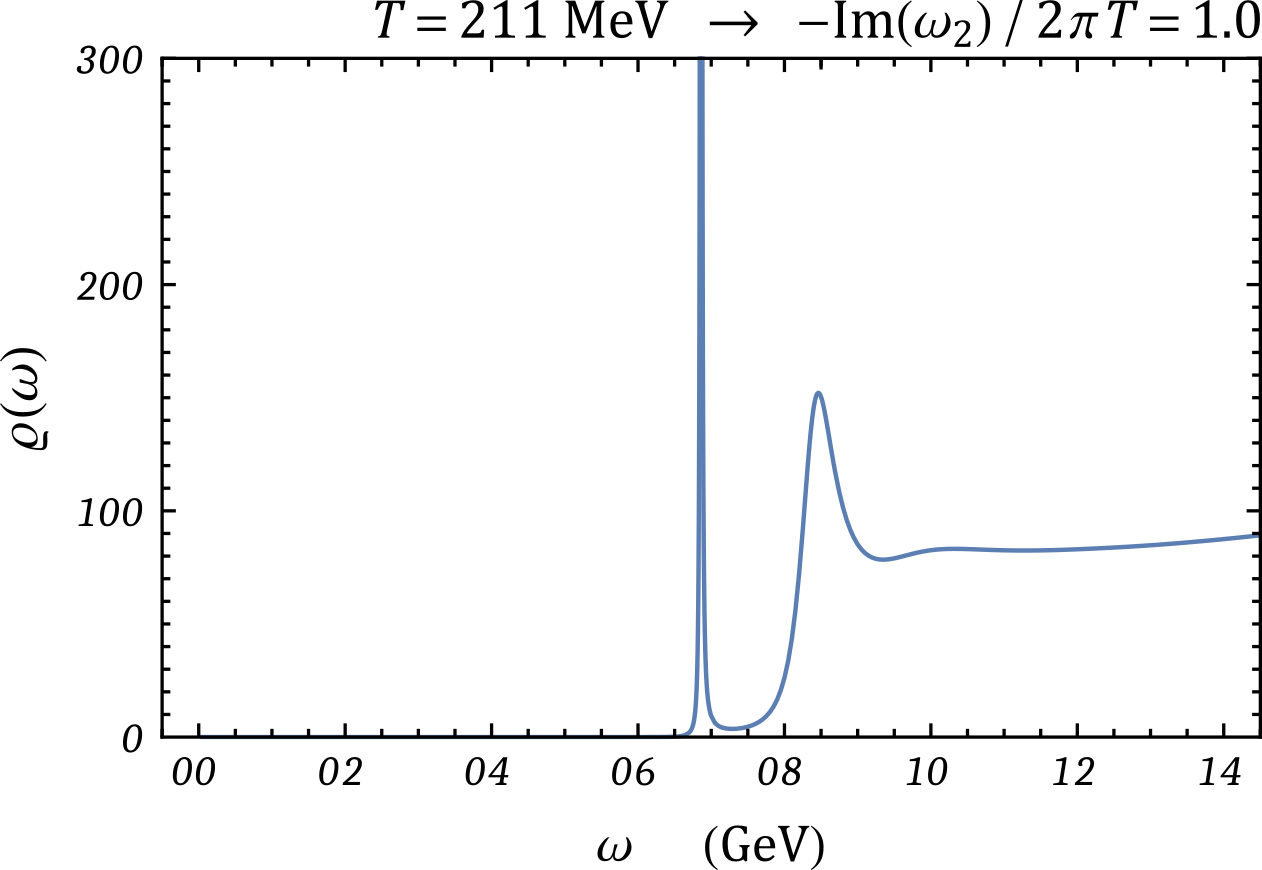}
    \\[10pt]
    \includegraphics[scale=.46]{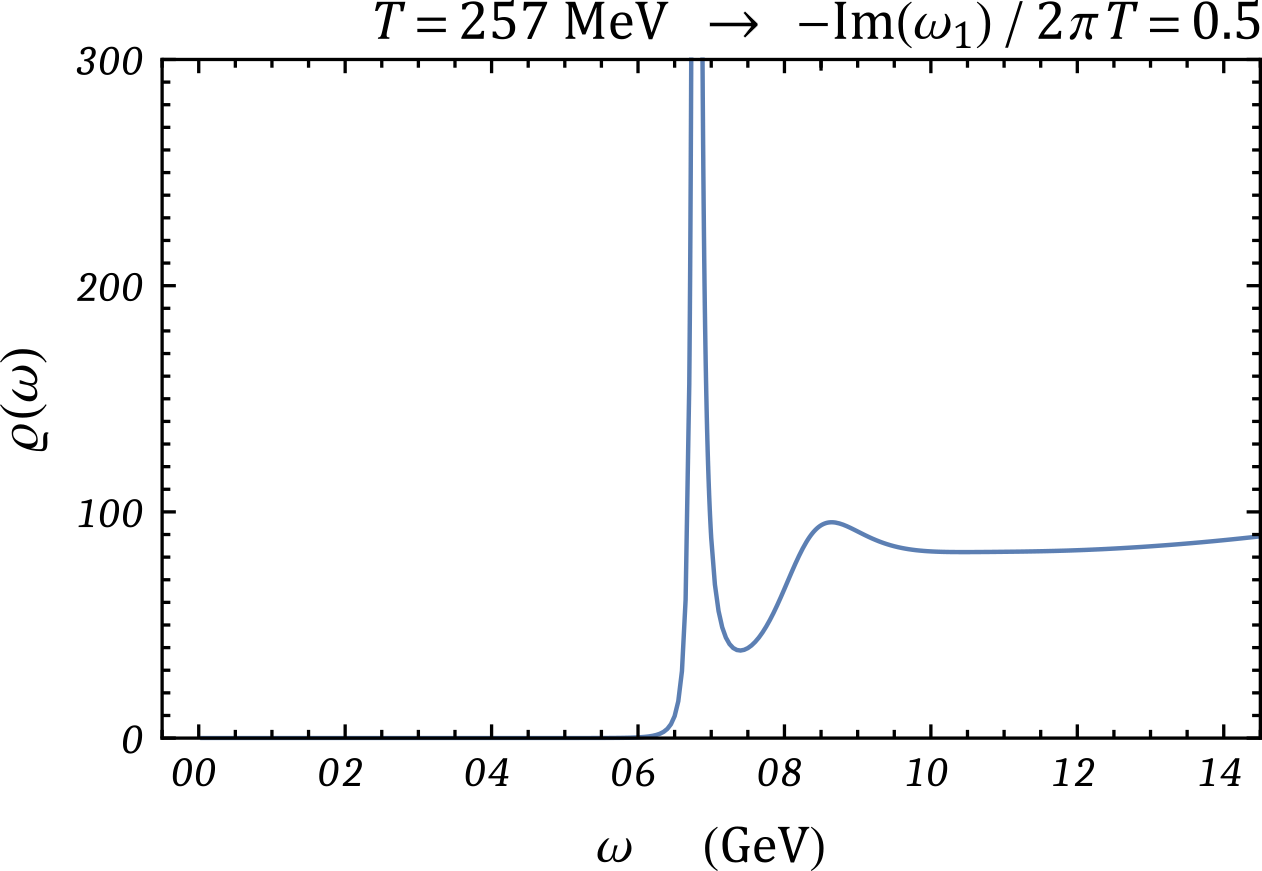} \quad
    \includegraphics[scale=.46]{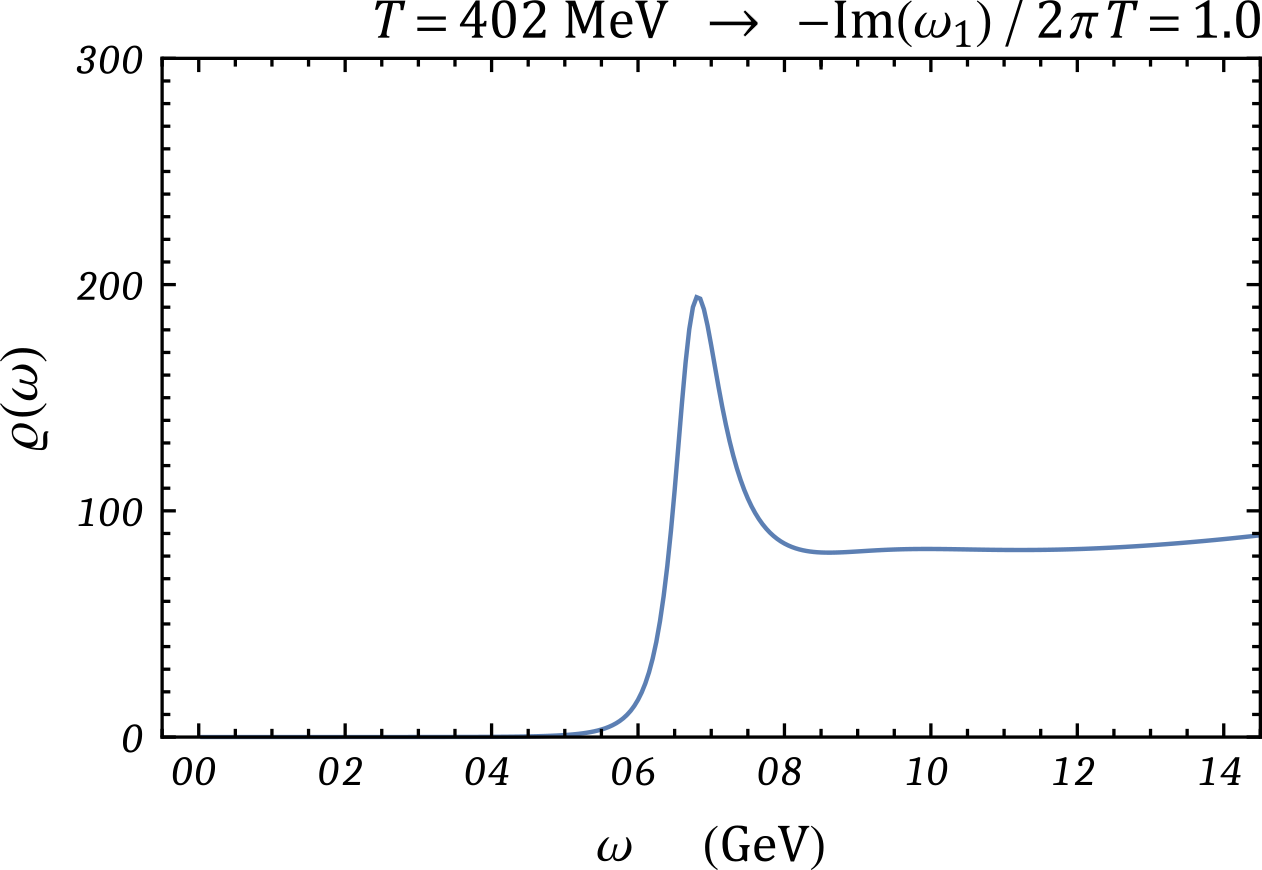}
    \\
    \caption{Spectral function for temperatures where $- \Im(ω_n)/2πT = 1.0 $ or $0.5$ for the different excitation levels $n$.}
    \label{fig: SFs}
\end{figure}

In order to confirm that the critical temperatures indeed correspond to values of $ T$ where the meson quasistates melt, we plot in figure \ref{fig: SFs} the spectral function for the temperatures where $\,-\Im(ω)/2πT$ is equal to 0.5 or 1.0 for the modes $ n = 0,1,2,3$. These spectral functions are calculated following the lines described in detail in, for example, reference \cite{Braga:2017oqw}. One notes that at $T = \SI{139}{\mega\eV}$ the first peaks, corresponding to $ n= 0,1,2$ are very sharp, corresponding to the presence of the quasistates in the thermal medium, while there is a very small ripple in the position of the mode $n =3$ indicating the melting of this quasisate. At $T = \SI{159}{\mega\eV}$ one notices that there is no more trace of the $n =3$ quasistate. In the other values of temperature present in figure \ref{fig: SFs}, we repeat the same analysis, but for the modes $n = 2$ and $n = 1$. The result found is similar. So, the critical temperatures that we found, where the DCE obtained from the nonregularized energy density can indeed be associated to a melting temperature.

% ------------------------------------------

\begin{figure}[htb!]
    \centering
    \includegraphics[scale=.41]{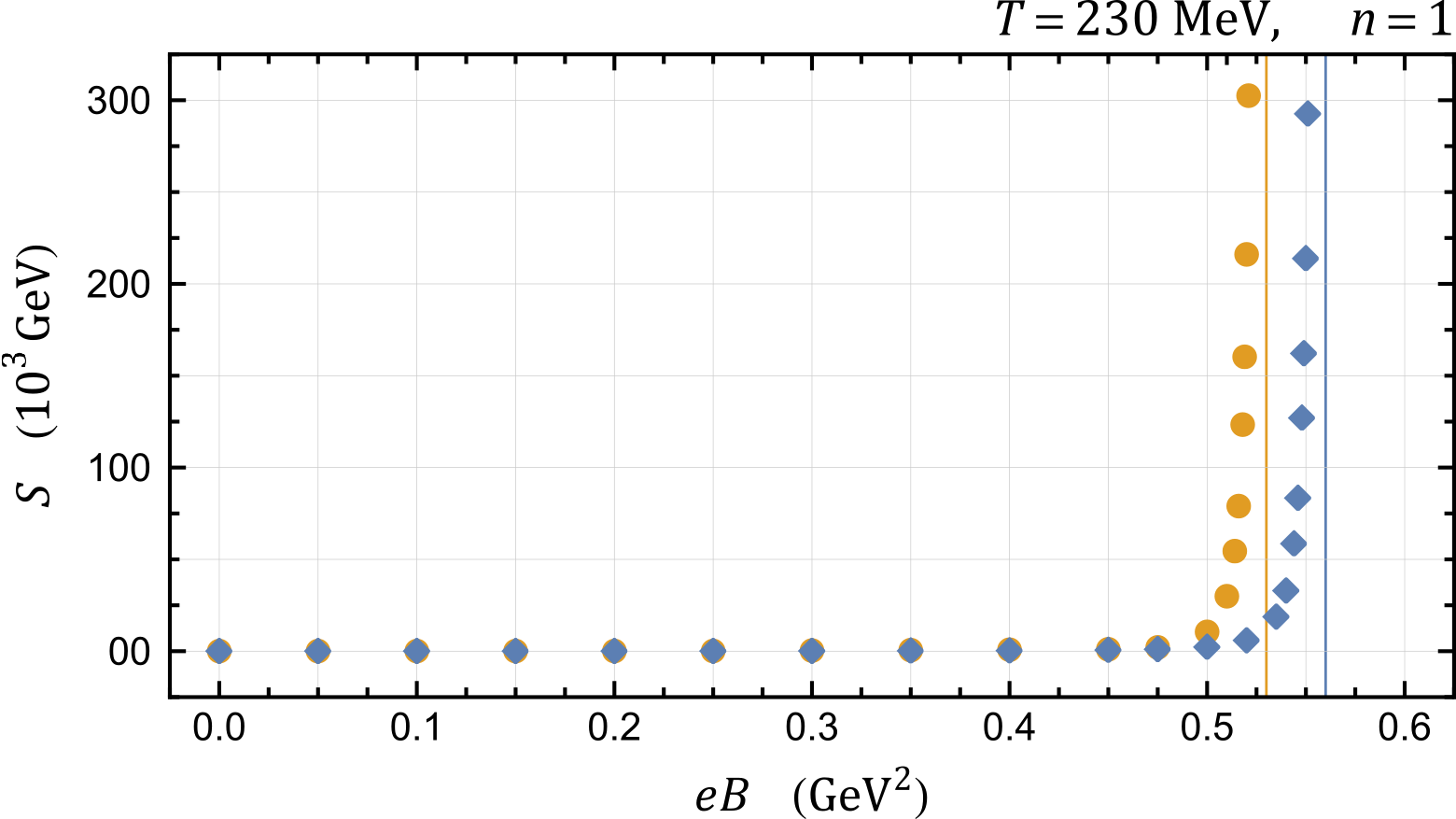}\\[8pt]
    \raisebox{2pt}{\includegraphics[scale=.36]{figs/MarkDiamond.png}} \quad Transverse Polarization\\[3pt]
    \raisebox{2pt}{\includegraphics[scale=.36]{figs/MarkDisk.png}}    \quad Longitudinal Polarization
    \caption{DCE as a function of the magnetic field $eB$ for $ T = \SI{230}{\mega\eV}$. Vertical lines at $eB = \SI{0.53}{\giga\eV^2}$ and $eB = \SI{0.56}{\giga\eV^2}$}
    \label{fig: DCEB}
\end{figure}

Now, in order to complete our analysis, let us consider the case with magnetic field in order to understand how can the DCE tell us when a meson dissociates by the effect of the magnetic field. With this purpose we plot in figure \ref{fig: DCEB} the variation of the DCE with the magnetic field $eB$ for $ T = \SI{230}{\mega\eV}$, with the same convention of colors for the transverse and longitudinal cases adopted in section \ref{sec: CE}. The vertical lines indicate the values of the fields where the DCE becomes singular, corresponding to the melting of the heavy vector meson quasistate.

\begin{figure}[htb!]
    \centering
    \includegraphics[scale=.45]{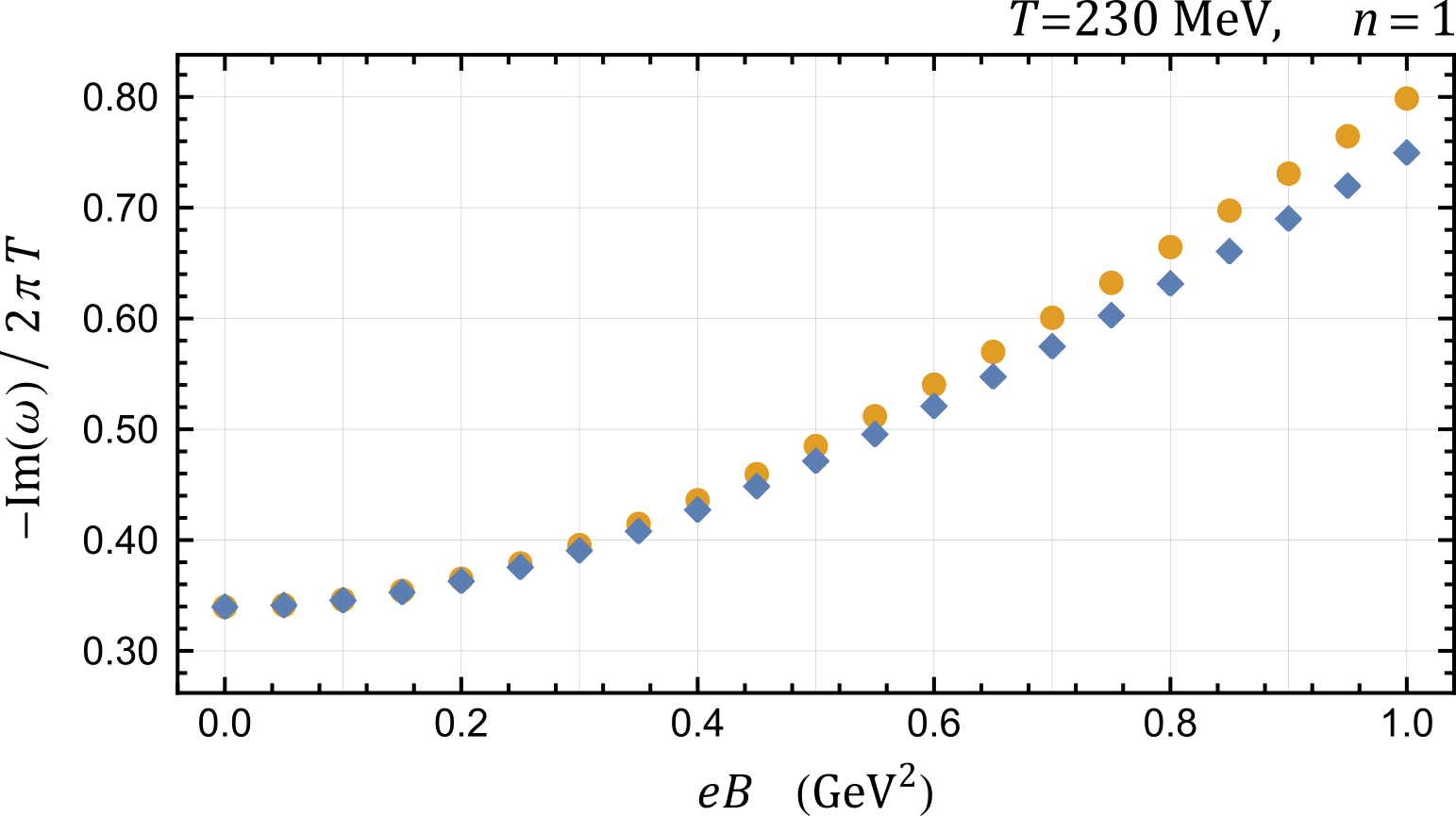}\\[8pt]
    \raisebox{2pt}{\includegraphics[scale=.36]{figs/MarkDiamond.png}} \quad Transverse Polarization\\[3pt]
    \raisebox{2pt}{\includegraphics[scale=.36]{figs/MarkDisk.png}}    \quad Longitudinal Polarization
    \caption{Value of the factor  $\,-\Im(ω)/2πT$ for $n = 1$ as a functions of the magnetic field $eB$.}
    \label{fig: factorB}
\end{figure}

Then, in figure \ref{fig: factorB} we show the variation of the factor $ - \Im(ω)/2πT$ as a function of $eB$. One notices that as in the case of the variation with $T$, this is the factor that controls the divergence of the DCE and the corresponding representation of the complete dissociation process.

% ----------------------------------------------------------
\vspace{.75\baselineskip}
\section{Conclusions}
\label{sec: Conclusions}

We presented in this paper the results of the numerical computations of the quasinormal modes associated with the four lowest radial excitation levels of bottomonium in a plasma with a constant background magnetic field $eB$.

Using these solutions we calculated the DCE for these states. In order to find a result that is finite for all the values of $n$, $T$ and $eB$ we introduced a regularized energy density that is a square integrable function of the coordinate $z$. The result obtained shows that the DCE increases with the radial excitation level and also with the value of the field $eB$. As discussed in the introduction, for many different systems it has been observed that the DCE works as an indicator of stability. The more stable the system is, the lower the value of the DCE. Therefore, our results are consistent with the fact that the higher excited states of bottomonium are subject to a stronger dissociation effect in the plasma, so that they are more unstable.

On the other hand, it is known from the calculation of spectral functions that the presence of magnetic fields enhances the dissociation effect of bottomonium in a plasma \cite{Braga:2019yeh}. So that, by increasing the $eB$ field it is expected that the bottomonium quasistates become more unstable. This is consistent with our finding that the DCE increases with the magnetic field.

The same kind of behavior was inferred from the analysis of the imaginary part of the quasinormal frequencies $\Im(ω)$. As shown in figure \ref{fig: QNMs2}, $|\Im(ω)|$ increases with the value of the excitation level $n$ and with the field $eB$. An increase in this quantity is associated with an enhancement in the dissociation in the medium and, thus, with instability. However, the increase in $|\Im(ω)|$ with the field obtained here is stronger for the larger values of $n$. In contrast, for the DCE the effect of the variation with the magnetic field is approximately the same for all values of $n$. This fact can be seen as an indication that the DCE should not be taken as a quantitative measure of the dissociation. We mean: we confirmed for the bottomonium excited states in a magnetic field background that the higher is the DCE, the more unstable the system. However, we noticed that the rate of increase in the DCE should not be taken as a rate of increase in the degree of instability.

Finally we explored the possibility of using the energy density itself in order to calculate the DCE, without the regularization process. Surprisingly, it emerged the result that the DCE becomes singular at some values of $n$, $T$ and $eB$. We found out that this singular behavior can be associated with the complete dissociation (melting) of the quasiparticles. It is natural to associate a situation of maximum instability to the behavior $S \to \infty$. So the use of the nonregularized energy density leads one to a DCE that is capable of identifying the process of complete dissociation of the vector mesons.

For some alternative interesting approaches to quarkonium in a thermal medium, see for example \cite{Dudal:2014jfa, Dudal:2018rki}.

\noindent {\bf Acknowledgments:} N.R.F.B. is partially supported by CNPq --- Conselho Nacional de Desenvolvimento Científico e Tecnologico grant 307641/2015-5 and by FAPERJ --- Fundação Carlos Chagas Filho de Amparo à Pesquisa do Estado do Rio de Janeiro. The authors received also support from Coordenação de Aperfeiçoamento de Pessoal de Nível Superior --- Brasil (CAPES), Finance Code 001.

% ==========================================================
\setstretch{1.2}
\bibliography{bibliography}
% ==========================================================

\end{document}